\documentclass[12pt]{article}
\usepackage[left=2.5cm,top=2.50cm,right=2.5cm,bottom=2.50cm]{geometry}
\usepackage{mathrsfs}
\usepackage{amsmath,amssymb,latexsym,color,cancel,graphicx,bbm,colortbl}
\usepackage[english]{babel}
\usepackage[latin1]{inputenc}
\usepackage{ragged2e}
\usepackage{float}
\usepackage{cite}
\usepackage{graphicx}
\usepackage{placeins}
\usepackage{nccmath}
\usepackage{hyperref}
\usepackage[toc,title]{appendix}
\usepackage{url}
\DeclareMathAlphabet{\mathpzc}{OT1}{pzc}{m}{it}
\begin{document}
\date{}
\title{$\mathfrak{su}(1,1)$ Symmetry and Exact Solutions of the
Dunkl--Klein--Gordon Equation in Higher Dimensions}
 \maketitle
\begin{center}
{\large M. Salazar-Ram\'{i}rez$^{1}$}\\[0.3em]
\begin{minipage}{0.9\textwidth}
\centering
\small Escuela Superior de C\'omputo, Instituto Polit\'ecnico Nacional,\\
Av. Juan de Dios B\'atiz esq. Av. Miguel Oth\'on de Mendiz\'abal, Col. Lindavista,\\
Alc. Gustavo A. Madero, C.P. 07738, Ciudad de M\'exico, M\'exico.
\end{minipage}
\footnotetext[1]{msalazarra@ipn.mx (Corresponding author)}
\end{center}

\vspace{0.7em}
\begin{center}
{\large B. C. L\"utf\"uo\u{g}lu$^{2}$}\\[0.3em]

\begin{minipage}{0.9\textwidth}
\centering
\small Department of Physics, University of Hradec Kr\'alov\'e,\\
Rokitansk\'eho 62, 500 03 Hradec Kr\'alov\'e, Czechia.
\end{minipage}
\footnotetext[2]{bekir.lutfuoglu@uhk.cz}
\end{center}
\vspace{0.7em}

\begin{center}
{\large J. A. Mart\'{i}nez-Nu\~no$^{3}$}\\[0.3em]
\begin{minipage}{0.9\textwidth}
\centering
\small Escuela Superior de C\'omputo, Instituto Polit\'ecnico Nacional,\\
Av. Juan de Dios B\'atiz esq. Av. Miguel Oth\'on de Mendiz\'abal, Col. Lindavista,\\
Alc. Gustavo A. Madero, C.P. 07738, Ciudad de M\'exico, M\'exico.
\end{minipage}
\footnotetext[3]{jmartinezn@ipn.mx}
\end{center}

\begin{abstract}
We investigate the $d$-dimensional Dunkl--Klein--Gordon equation for a scalar particle within an algebraic framework. By employing Schr\"odinger factorization, we construct the generators of the $\mathfrak{su}(1,1)$ algebra and establish the associated symmetry of the radial sector. The corresponding energy spectra are derived from unitary irreducible representations, and the Sturmian radial bases are obtained analytically. We analyze the $d$-dimensional Dunkl--Klein--Gordon oscillator and the bound-state sector of the $d$-dimensional Dunkl--Klein--Gordon equation with a Dunkl--Coulomb-like potential. Furthermore, $\mathrm{SU}(1,1)$ coherent states are constructed and their time evolution is analyzed, revealing a characteristic radial oscillation behavior. The results show that the Dunkl deformation modifies the spatial and spectral structure of the system and, in the oscillator sector, introduces explicit parity-dependent energy shifts while preserving the underlying radial algebraic dynamics.
\end{abstract}


\section{Introduction}

Relativistic wave equations constitute one of the cornerstones of modern theoretical physics, providing the fundamental framework for describing elementary particles and their interactions. Among them, the Klein--Gordon equation occupies a distinguished position as the relativistic wave equation for spin-zero particles. Beyond its role in particle physics, the Klein--Gordon equation has served as an important theoretical laboratory for investigating bound-state problems, symmetry structures, and exactly solvable models in relativistic quantum mechanics. Owing to its broad applicability, it has been extensively employed in nuclear, atomic, and molecular physics, as well as in effective descriptions arising from quantum field theory and gravitational models.

In particular, the extension of relativistic quantum systems to arbitrary spatial dimensions has attracted considerable attention because higher-dimensional formulations naturally emerge in Kaluza--Klein theories, string-inspired scenarios, and higher-dimensional gravity \cite{Appelquist1987,Overduin1997,Green1987}. Moreover, the study of relativistic systems in arbitrary dimensions often reveals hidden mathematical structures and nontrivial spectral properties that remain obscured in the conventional three-dimensional setting.

Consequently, numerous investigations have been devoted to exact and approximate solutions of the $d$-dimensional Klein--Gordon equation for a variety of interaction potentials, including Coulomb, Morse, Manning--Rosen, Kratzer, and exponential-type interactions, revealing interdimensional degeneracies and dimension-dependent spectral effects \cite{Dong2003,Chen2014,Xie2015,Ikot2016}.

The search for exact solutions of relativistic wave equations has stimulated the development of powerful analytical and algebraic techniques. Among them, factorization methods occupy a prominent position because they establish a direct connection between differential equations and symmetry generators. Such approaches allow the spectral problem to be reformulated in terms of representation theory, often leading to exact energy spectra and eigenfunctions without explicitly solving the corresponding differential equations.

Since the pioneering works of Schr\"odinger, Infeld, and Hull, factorization techniques have become an indispensable tool in the study of exactly solvable quantum systems \cite{Infeld1951}. In many physically relevant cases, the underlying dynamical symmetry is associated with noncompact Lie algebras whose representation spaces naturally encode the bound-state spectrum \cite{Miller1968,Wybourne1974}.

Among these structures, the Lie algebra $\mathfrak{su}(1,1)$ plays a particularly important role. It appears in a wide variety of physical systems, including harmonic oscillators, Coulomb-like interactions, quantum optical models, molecular systems, and relativistic wave equations. Furthermore, the associated unitary irreducible representations provide a natural framework for constructing coherent states and studying quantum dynamics from a group-theoretical perspective \cite{Barut1971,Perelomov1986}.

A different but closely related direction emerged from Wigner's observation that the equations of motion do not uniquely determine the canonical commutation relations of quantum mechanics \cite{Wigner1950}. Motivated by this idea, Yang introduced a generalized realization of the oscillator algebra involving reflection operators, thereby opening the possibility of extending conventional quantum mechanics through differential--difference constructions \cite{Yang1951}. These developments culminated in the work of Dunkl, who introduced a family of differential--difference operators associated with finite reflection groups \cite{Dunkl1989,Dunkl1991}.

Since then, Dunkl operators have attracted considerable attention because of their deep connections with harmonic analysis, special functions, and representation theory \cite{Rosler2003}. Furthermore, they play a central role in the study of integrable many-body systems, particularly those related to Calogero--Sutherland models \cite{Calogero1969,Sutherland1971}. From a physical perspective, the inclusion of reflection operators generates parity-dependent deformations that modify the spectral and dynamical properties of quantum systems while preserving exact solvability in many important cases. Such deformations also arise naturally in generalized oscillator algebras, supersymmetric extensions, and Wigner--Dunkl quantum mechanics \cite{Plyushchay1996,Plyushchay2000,Genest2013a,Genest2014,Junker2022, ChungJunkerDongHassanabadi2023}.

The resulting framework, commonly referred to as Dunkl quantum mechanics, has revealed hidden symmetries, superintegrable structures, and generalized dynamical algebras in oscillator and Coulomb systems with reflection operators \cite{Genest2013a,Genest2014,Genest2013b,Ghazouani2020}. Its scope has subsequently been extended to coherent states, supersymmetric models, and higher-dimensional systems, revealing nontrivial effects of the Dunkl deformation
on the spectral and quantum properties of these systems ~\cite{Junker2022,Chung2023,Sedaghatnia2023,HamilLutfuogluMerad2025a,Ghazouani2021}.

Dunkl operators have also been incorporated into relativistic wave equations. Exact solutions of the Dunkl--Klein--Gordon oscillator and Coulomb problems have been obtained in two, three, and arbitrary spatial
dimensions \cite{Mota2021A,Hamil2022,HamilLutfuogluMerad2025b}. Related investigations have addressed Dirac--Dunkl systems, generalized relativistic oscillators, inverse-power interactions, and path-integral
formulations \cite{OjedaGuillen2020,Salazar2018,SchulzeHalberg2023KG, Rouabhia2023,Bouguerne2024,Benzair2024a,Benzair2024b,Benzair2026}. The construction of coherent states and $\mathfrak{su}(1,1)$ realizations
has further demonstrated the usefulness of representation-theoretical methods in relativistic Dunkl systems~\cite{Salazar2026}.

Recent studies have addressed the Dunkl--Klein--Gordon equation in higher dimensions using different approaches. Hamil et al.~\cite{HamilLutfuogluMerad2025b} obtained exact analytical solutions for the $d$-dimensional oscillator and Coulomb problems, expressing the energy spectra and eigenfunctions in terms of confluent hypergeometric functions and additionally analyzing the scattering states in the Coulomb sector. From a different perspective, Benzair et al.~\cite{Benzair2024b,Benzair2026} developed path-integral formulations for the Klein--Gordon--Dunkl oscillator and, subsequently, for the Coulomb--Dunkl--Klein--Gordon problem in higher dimensions, obtaining exact propagators, energy spectra, and the corresponding wave functions.

The present work complements these studies by focusing on the underlying algebraic structure and developing a unified $\mathfrak{su}(1,1)$ treatment of both the oscillator and Coulomb sectors in arbitrary dimensions. Within this framework, Schr\"odinger factorization is employed to construct the radial generators explicitly, identify the positive-discrete representations, and derive the Sturmian bases algebraically. The same representation-theoretical structure also allows us to construct the corresponding Perelomov coherent states and study their temporal evolution.

The $d$-dimensional Dunkl--Klein--Gordon formulation is not merely a formal extension of the three-dimensional problem. It provides a unified framework for analyzing how dimensionality, angular momentum, and reflection structure
modify the dynamics of a relativistic scalar particle. The Dunkl parameters characterize the reflection-induced deformation and organize the state space into sectors of definite parity. Consequently, they modify the effective
angular interaction, energy spectrum, and radial localization while preserving an exactly tractable structure \cite{Dunkl1989,Rosler2003,Genest2013a}.

The spatial dimension, in turn, introduces a geometric contribution that makes it possible to distinguish ordinary dimensional effects from those induced by the deformation. Moreover, when all Dunkl parameters vanish, the conventional Klein--Gordon theory is continuously recovered. The model thus allows the spectral and dynamical properties arising specifically from the differential--reflection structure to be identified.

The purpose of the present work is to develop a unified algebraic treatment of the $d$-dimensional Dunkl--Klein--Gordon equation based on $\mathfrak{su}(1,1)$ symmetry. By employing Schr\"odinger factorization, we construct the generators of the algebra and establish the associated dynamical symmetry of the radial sector in arbitrary spatial dimensions. This approach allows the energy spectrum and the corresponding Sturmian basis to be obtained directly from the unitary irreducible representations of $\mathfrak{su}(1,1)$.

As applications of the formalism, we investigate both the $d$-dimensional Dunkl--Klein--Gordon oscillator and the bound-state sector of a Dunkl--Coulomb-like interaction within the same algebraic formulation. We further construct the associated $\mathrm{SU}(1,1)$ Perelomov coherent states and analyze their time evolution. The resulting description provides a systematic higher-dimensional extension of previous relativistic Dunkl models and unifies symmetry, energy spectra, Sturmian functions, and coherent-state dynamics through a common representation-theoretical structure.

The paper is organized as follows. In Sec.~2, we present the $d$-dimensional Dunkl--Klein--Gordon equation and derive its radial form in hyperspherical coordinates. In Sec.~3, we develop the $\mathfrak{su}(1,1)$ algebraic formulation of the Dunkl--Klein--Gordon oscillator and obtain its energy spectrum and Sturmian basis. In Sec.~4, the associated $\mathrm{SU}(1,1)$ coherent states are constructed, while their time evolution is analyzed in Sec.~5. Section~6 is devoted to the bound-state sector of the Dunkl--Klein--Gordon equation with a Coulomb-like interaction. In Sec.~7, we discuss the physical framework and consistency checks, including the Dunkl measure, normalization and regularity conditions, symmetry properties, and relevant limiting cases. Finally, Sec.~8 presents the concluding remarks.

\section{The d-Dimensional Dunkl--Klein--Gordon Equation}
The Dunkl--Klein--Gordon equation in $d$ dimensions is written as \cite{HamilLutfuogluMerad2025b}
\begin{equation}
\left(E^{2}+D_{j}^{2}-m^{2}\right)\Psi(x)=0,
\end{equation}
where $\mathbf{x}=(x_1,x_2,\ldots,x_d)$, $D_j=\partial_{x_j}+(\mu_j/x_j)(1-R_j)$ denotes the Dunkl derivative,
and summation over repeated spatial indices is understood. The hyperspherical coordinates are introduced as
\begin{align}\nonumber
x_{1} &= r\cos\theta_{1}\sin\theta_{2}\sin\theta_{3}\cdots\sin\theta_{d-1}, \\\nonumber
x_{2} &= r\sin\theta_{1}\sin\theta_{2}\sin\theta_{3}\cdots\sin\theta_{d-1}, \\\nonumber
x_{3} &= r\cos\theta_{2}\sin\theta_{3}\sin\theta_{4}\cdots\sin\theta_{d-1}, \\\nonumber
&\vdots \nonumber \\\nonumber
x_{j} &= r\cos\theta_{j-1}\sin\theta_{j}\sin\theta_{j+1}\cdots\sin\theta_{d-1},
\qquad 3\leq j \leq d-1,\\\nonumber
&\vdots \nonumber \\
x_{d} &= r\cos\theta_{d-1},
\end{align}
with $r\in(0,\infty)$, $0\leq \theta_{1}\leq 2\pi$, and
$0\leq \theta_{k}\leq \pi$ for $k=2,\dots,d-1$.
These coordinates satisfy
\begin{equation}
x_{1}^{2}+x_{2}^{2}+\cdots+x_{d}^{2}=r^{2}.
\end{equation}
In hyperspherical coordinates, the Laplacian operator takes the form\cite{HamilLutfuogluMerad2025b}
\begin{align}
\Delta &=
\frac{\partial^{2}}{\partial r^{2}}
+
\frac{d-1}{r}\frac{\partial}{\partial r}
+
\frac{1}{r^{2}}
\sum_{j=1}^{d-2}
\frac{1}{\sin^{2}\theta_{j+1}\sin^{2}\theta_{j+2}\cdots\sin^{2}\theta_{d-1}}
\left(
\frac{\partial^{2}}{\partial \theta_{j}^{2}}
+
(j-1)\tan\theta_{j}\frac{\partial}{\partial \theta_{j}}
\right)
\nonumber\\
&\quad
+
\frac{1}{r^{2}}
\left[
\frac{1}{\sin^{d-2}\theta_{d-1}}
\frac{\partial}{\partial \theta_{d-1}}
\left(
\sin^{d-2}\theta_{d-1}
\frac{\partial}{\partial \theta_{d-1}}
\right)
\right].
\end{align}
Moreover, the volume element is given by
\begin{equation}
\prod_{j=1}^{d}dx_{j}
=
r^{d-1}\,dr
\prod_{j=1}^{d-1}
(\sin\theta_{j})^{\,j-1}d\theta_{j}.
\end{equation}
After expressing the Dunkl derivatives in hyperspherical coordinates, the
$d$-dimensional Dunkl--Klein--Gordon equation can be written as\cite{HamilLutfuogluMerad2025b}
\begin{equation}
\left[
A_{r}
+
\frac{J_{\theta_{1}}}{r^{2}\sin^{2}\theta_{2}\sin^{2}\theta_{3}\cdots\sin^{2}\theta_{d-1}}
+
\frac{J_{\theta_{2}}}{r^{2}\sin^{2}\theta_{3}\cdots\sin^{2}\theta_{d-1}}
+\cdots+
\frac{1}{r^{2}}J_{\theta_{d-1}}
\right]\Psi(x)=0,
\end{equation}
where
\begin{equation}
A_{r}=
\frac{\partial^{2}}{\partial r^{2}}
+
\frac{d-1+2(\mu_{1}+\mu_{2}+\cdots+\mu_{d})}{r}\frac{\partial}{\partial r}
+
E^{2}-m^{2}.
\end{equation}
The angular operators $J_{\theta_{1}},J_{\theta_{2}},\dots,J_{\theta_{d-1}}$ depend only on the angular variables. Their first members are
\begin{align}
J_{\theta_{1}}&=
-\frac{\partial^{2}}{\partial \theta_{1}^{2}}
+
2\left(\mu_{1}\tan\theta_{1}-\mu_{2}\cot\theta_{1}\right)
\frac{\partial}{\partial \theta_{1}}
+
\frac{\mu_{1}}{\cos^{2}\theta_{1}}(1-R_{1})
+
\frac{\mu_{2}}{\sin^{2}\theta_{1}}(1-R_{2}),
\\[1ex]
J_{\theta_{2}}&=
-\frac{\partial^{2}}{\partial \theta_{2}^{2}}
-
\left[
\bigl(1+2(\mu_{1}+\mu_{2})\bigr)\cot\theta_{2}
-
2\mu_{3}\tan\theta_{2}
\right]
\frac{\partial}{\partial \theta_{2}}
+
\frac{\mu_{3}}{\cos^{2}\theta_{2}}(1-R_{3}),
\end{align}
and so on up to $J_{\theta_{d-1}}$. The corresponding reflection operators act on the angular variables according to
\begin{equation}
\begin{aligned}
R_1 f(r,\theta_1,\theta_2,\ldots,\theta_j,\ldots,\theta_{d-1})
&=
f(r,\pi-\theta_1,\theta_2,\ldots,\theta_j,\ldots,\theta_{d-1}),
\\
R_2 f(r,\theta_1,\theta_2,\ldots,\theta_j,\ldots,\theta_{d-1})
&=
f(r,-\theta_1,\theta_2,\ldots,\theta_j,\ldots,\theta_{d-1}),
\\
&\hspace{4cm}\vdots
\\
R_j f(r,\theta_1,\theta_2,\ldots,\theta_j,\ldots,\theta_{d-1})
&=
f(r,\theta_1,\theta_2,\ldots,\pi-\theta_j,\ldots,\theta_{d-1}),
\\
&\hspace{4cm}\vdots
\\
R_d f(r,\theta_1,\theta_2,\ldots,\theta_j,\ldots,\theta_{d-1})
&=
f(r,\theta_1,\theta_2,\ldots,\theta_j,\ldots,\pi-\theta_{d-1}).
\end{aligned}
\label{reflection-operators-angular}
\end{equation}
while the first angular equations read
\begin{align}
\left(J_{\theta_{1}}+\lambda_{1}^{2}\right)\Theta_{1}(\theta_{1})&=0,
\\
\left(
J_{\theta_{2}}+\frac{\lambda_{1}^{2}}{\sin^{2}\theta_{2}}+\lambda_{2}^{2}
\right)\Theta_{2}(\theta_{2})&=0,
\\
&\vdots \nonumber\\
\left(
J_{\theta_{d-1}}+\frac{\lambda_{d-2}^{2}}{\sin^{2}\theta_{d-1}}+\varpi^{2}
\right)\Theta_{d-1}(\theta_{d-1})&=0.
\end{align}
Here, $\lambda_{1},\lambda_{2},\dots,\lambda_{d-2}$ and $\varpi$ are separation
constants. In particular, the radial separation constant $\varpi^{2}$ is
\begin{equation}\label{OMEGAL}
\varpi^{2}=4L\left(L+\sum_{i=1}^d\mu_i+\frac{d-2}{2}\right),
\end{equation}
where $L\equiv\sum_{j=1}^{d-1}\ell_j$ is not an additional independent quantum number, but a compact notation for the sum of the angular quantum numbers arising from the hyperspherical separation. In the non-deformed limit, the conventional $d$-dimensional orbital quantum number $\ell$ is related to the present notation by $\ell=2L$. The separated wavefunctions are chosen in definite reflection sectors, such that $R_i\Psi=s_i\Psi$, where $s_i=\pm1$ for $i=1,\ldots,d$.

Introducing the Dunkl oscillator interaction, the Klein--Gordon equation takes the form
\begin{equation}
\left[
E^{2}
-
\left(
\frac{1}{i}D_{j}+im\omega x_{j}
\right)
\left(
\frac{1}{i}D_{j}-im\omega x_{j}
\right)
-m^{2}
\right]\Psi(x)=0.
\end{equation}
Therefore, the radial equation of the Dunkl--Klein--Gordon oscillator in $d$ dimensions is
\begin{equation}\label{EDSEHD}
\begin{split}
&\left[\frac{d^{2}}{dr^{2}}+\frac{d-1+2(\mu_{1}+\mu_{2}+\cdots+\mu_{d})}{r}\frac{d}{dr}+2m\omega\left(\mu_{1}s_{1}+\mu_{2}s_{2}+\cdots+\mu_{d}s_{d}+\frac{d}{2}\right)\right. \\
&\quad \left. -\,m^{2}\omega^{2}r^{2}+E^{2}-m^{2}-\frac{\varpi^{2}}{r^{2}}\right]R(r)=0.
\end{split}
\end{equation}
\section{Schr\"odinger Factorization and $\mathfrak{su}(1,1)$ Algebraic Structure}
 Schr\"odinger factorization reveals the underlying algebraic structure of the system and provides a systematic framework for deriving both the energy spectrum and the corresponding eigenfunctions. Furthermore, this approach naturally leads to the construction of ladder operators and the identification of the associated $\mathfrak{su}(1,1)$ symmetry of the model. In order to eliminate the first-derivative term in Eq.~(\ref{EDSEHD}), we introduce the transformation
\begin{equation}\label{TRAF}
R(r)=
r^{-\frac{d-1+2(\mu_{1}+\mu_{2}+\cdots+\mu_{d})}{2}}\,F(r).
\end{equation}
Substituting this expression into the radial equation, we obtain
\begin{equation}
\begin{aligned}
\Bigg[ \frac{d^{2}}{dr^{2}} &- m^{2}\omega^{2}r^{2} - \frac{\varpi^{2} + (\mu_{1}+\cdots+\mu_{d})(\mu_{1}+\cdots+\mu_{d}+d-2) + \left(\frac{d}{2}-1\right)^{2}-\frac{1}{4}}{r^{2}} \\
&+ E^{2}-m^{2} + 2m\omega \left( \mu_{1}s_{1}+\cdots+\mu_{d}s_{d} +\frac{d}{2} \right) \Bigg] F(r) = 0,
\end{aligned}
\end{equation}
introducing the dimensionless radial variable
\begin{equation}\label{DIM}
\rho=\sqrt{m\omega}\,r,
\end{equation}
the radial equation becomes
\begin{equation}\label{EDSOPR}
\begin{aligned}
\Bigg[-\rho^2\frac{d^2}{d\rho^2}+\rho^4+\varpi^2
+(\mu_1+\cdots+\mu_d)(\mu_1+\cdots+\mu_d+d-2)
+\left(\frac d2-1\right)^2-\frac14
\\
-\frac{E^2-m^2+2m\omega
\left(\mu_1s_1+\cdots+\mu_ds_d+\frac d2\right)}
{m\omega}\rho^2\Bigg]F(\rho)=0.
\end{aligned}
\end{equation}
By means of the Schr\"odinger factorization framework, we can obtain a closed realization of the $\mathfrak{su}(1,1)$ generators with the following ansatz
\begin{equation}\label{FACSHD}
\left[\rho\frac{d}{d\rho}+\mathfrak{A}\rho^2+\mathfrak{B}\right]\biggl[-\rho\frac{d}{d\rho}+\mathfrak{C}\rho^2+\mathfrak{F}\biggr]F(\rho)=\mathfrak{G}F(\rho).
\end{equation}
Expanding the product of the differential operators in equation~(\ref{FACSHD}) and collecting the terms with equal powers of $\rho$, one obtains
\begin{equation}\label{EDSOHD2S}
\resizebox{0.98\linewidth}{!}{$\displaystyle
\left[
-\rho^{2}\frac{d^{2}}{d\rho^{2}}
-\rho\left((\mathfrak{A}-\mathfrak{C})\rho^{2}
+\mathfrak{B}+1-\mathfrak{F}\right)\frac{d}{d\rho}
+\mathfrak{A}\mathfrak{C}\rho^{4}
+\left((\mathfrak{B}+2)\mathfrak{C}
+\mathfrak{A}\mathfrak{F}\right)\rho^{2}
+\mathfrak{B}\mathfrak{F}
\right]F(\rho)=\mathfrak{G}F(\rho).
$}
\end{equation}
A direct term-by-term comparison between equation~(\ref{EDSOHD2S}) and the radial equation~(\ref{EDSOPR}), multiplied by $-\rho^{2}$, yields the following system of algebraic relations for the factorization parameters
\begin{equation}
\begin{gathered}
\mathfrak{A}-\mathfrak{C}=0,
\qquad
\mathfrak{B}+1-\mathfrak{F}=0,
\qquad
\mathfrak{A}\mathfrak{C}=1,
\\[1mm]
\mathfrak{G}-\mathfrak{B}\mathfrak{F}
=\Xi+\frac{1}{4},
\qquad
(\mathfrak{B}+2)\mathfrak{C}
+\mathfrak{A}\mathfrak{F}
=-2\Pi.
\end{gathered}
\end{equation}
The coefficients $\mathfrak{A}$, $\mathfrak{B}$, $\mathfrak{C}$, $\mathfrak{F}$, and $\mathfrak{G}$, implicitly defined in Eq.~(\ref{FACSHD}), are determined by expanding this equation and performing a term-by-term comparison. Their explicit expressions are listed in Table~1.
\begin{table}[H]
\centering
\caption{Schr\"odinger factorization parameters for the $d$-dimensional Dunkl--Klein--Gordon oscillator.}
\label{tab:ABCFGDKG}
\begin{tabular}{ccccc}
\hline\hline
$\mathfrak{A}$ & $\mathfrak{B}$ & $\mathfrak{C}$ & $\mathfrak{F}$ & $\mathfrak{G}$ \\
\hline
$+1$ & $-\Pi-\tfrac{3}{2}$ & $+1$ & $-\Pi-\tfrac{1}{2}$
& $\Xi+(\Pi+1)^{2}$ \\[1mm]
$-1$ & $+\Pi-\tfrac{3}{2}$ & $-1$ & $+\Pi-\tfrac{1}{2}$
& $\Xi+(\Pi-1)^{2}$ \\
\hline\hline
\end{tabular}
\end{table}
where
\begin{align}
\Pi &=
\frac{
E^{2}-m^{2}
+
2m\omega
\left(
\mu_{1}s_{1}+\mu_{2}s_{2}+\cdots+\mu_{d}s_{d}
+\frac{d}{2}
\right)
}{2m\omega}, \\\label{XI}
\Xi &=
\varpi^{2}
+
(\mu_{1}+\mu_{2}+\cdots+\mu_{d})
(\mu_{1}+\mu_{2}+\cdots+\mu_{d}+d-2)
+
\left(\frac{d}{2}-1\right)^{2}-\frac{1}{4}.
\end{align}
It then follows that the differential equation satisfied by $F(\rho)$ can be rewritten in a factorized form, constituting the basis of the algebraic treatment of the system within the Schr\"odinger factorization approach
\begin{equation}
\left(\mathcal{J}_{\mp}\mp 1\right)\mathcal{J}_{\pm}=\frac{1}{4}\left[\Xi+1\right],
\end{equation}
where
\begin{equation}
\mathcal{J}_{\pm}=\frac{1}{2}\left(\mp\rho\frac{d}{d\rho}+\rho^{2}-\frac{\Pi}{2}\pm\frac{1}{2}\right).
\end{equation}
These results naturally lead to the construction of two new operators, which play a central role in the algebraic description of the system
\begin{align}\label{OPSB1HD1}
\mathcal{Q}_{\pm}=\frac{1}{2}\left(\mp\rho\frac{d}{d\rho}+\rho^{2}\mp\frac{1}{2}-2\mathcal{K}_0\right),
\end{align}
where
\begin{equation}\label{TEROPC1HD1}
\mathcal{K}_0F(\rho)= \frac{1}{4}
\left[
-\frac{d^{2}}{d\rho^{2}}F(\rho)
+
\rho^{2}
+
\frac{\Xi}{\rho^{2}}
\right]F(\rho)=\frac{\Pi}{4}F(\rho).
\end{equation}
The connection between the factorization operators $\mathcal{J}_{\pm}$ and the generators $\mathcal{Q}_{\pm}$ follows from the eigenvalue equation $\mathcal{K}_0F(\rho)=(\Pi/4)F(\rho)$. Accordingly, the replacement $\Pi/2\rightarrow2K_0$ gives $\mathcal{Q}_{\pm}F(\rho)=J_{\pm}F(\rho)$.

Thus, $\mathcal{J}_{\pm}$ contain the spectral parameter $\Pi$ and act as factorization operators, whereas $\mathcal{Q}_{\pm}$ are their operator realization. The former do not define an independent $\mathfrak{su}(1,1)$ algebra; the closed realization is generated by $\mathcal{K}_0$ and $\mathcal{Q}_{\pm}$. Eqs.~(\ref{OPSB1HD1}) and (\ref{TEROPC1HD1}) are readily seen to define a closed realization of the $\mathfrak{su}(1,1)$ algebra
\begin{equation}\label{TEROP3}
[\mathcal{K}_0,\mathcal{Q}_{\pm}]=\pm\mathcal{Q}_{\pm}, \hspace{0.5cm} [\mathcal{Q}_{-}, \mathcal{Q}_{+}]=2\mathcal{K}_0.
\end{equation}
The operator conventions and the positive-discrete representation adopted here are summarized in Appendix A. To derive the energy spectrum of the higher-dimensional Dunkl--Klein--Gordon equation, we employ the unitary irreducible representations of the noncompact Lie algebra $\mathfrak{su}(1,1)$~\cite{Barut1971,Perelomov1986}.

\begin{equation}\label{OPTLA}
\mathcal{K}_0|k, n\rangle=(k+n) |k,n\rangle.
\end{equation}
Furthermore, the quadratic Casimir operator $\mathbb{C}^2$ obeys the following eigenvalue relation, which characterizes the irreducible representations of the algebra
\begin{equation}\label{CASC1}\nonumber
\mathbb{C}^2=\mathcal{K}_0(\mathcal{K}_0-1)-\mathcal{Q}_{+}\mathcal{Q}_{-}=\frac{\Xi}{4}-\frac{3}{16}=k(k-1),
\end{equation}
the above equation determines the Bargmann index $k$ associated with the present problem in the form
\begin{equation}\label{INBAR}
k=\frac{1}{2}
+
\frac{\sqrt{\Xi+\frac{1}{4}}}{2},
\end{equation}
the quantization condition follows immediately from the combined use of Eqs.~(\ref{TEROPC1HD1}), (\ref{OPTLA}) and (\ref{INBAR})
\begin{equation}
E^{2}-m^{2}
+
2m\omega
\left(
\mu_{1}s_{1}+\mu_{2}s_{2}+\cdots+\mu_{d}s_{d}
+\frac{d}{2}
\right)
=
4m\omega(k+n),
\end{equation}
from which the energy spectrum of the d-dimensional Dunkl--Klein--Gordon oscillator is obtained
\begin{equation}
E_{n,s_1,\dots,s_d}=\pm\sqrt{2m\omega\left[2n+1+\sqrt{\Xi+\frac{1}{4}}-S\right]+m^2},
\end{equation}
with $S=\mu_1 s_1+\mu_2 s_2+\cdots+\mu_d s_d+\frac{d}{2}$ and Eq.~(\ref{XI}), the previous expression can be rewritten as
\begin{equation}\label{SPECDK}
\medmath{E_{n,s_1,\ldots,s_d}
=
\pm\sqrt{
2m\omega\left[
2\left(n+\ell_1+\ell_2+\cdots+\ell_{d-1}\right)
+\mu_1+\mu_2+\cdots+\mu_d
-\left(\mu_1 s_1+\mu_2 s_2+\cdots+\mu_d s_d\right)
\right]
+m^2
}}.
\end{equation}

\begin{figure}[H]
    \centering
    \includegraphics[width=0.98\textwidth]{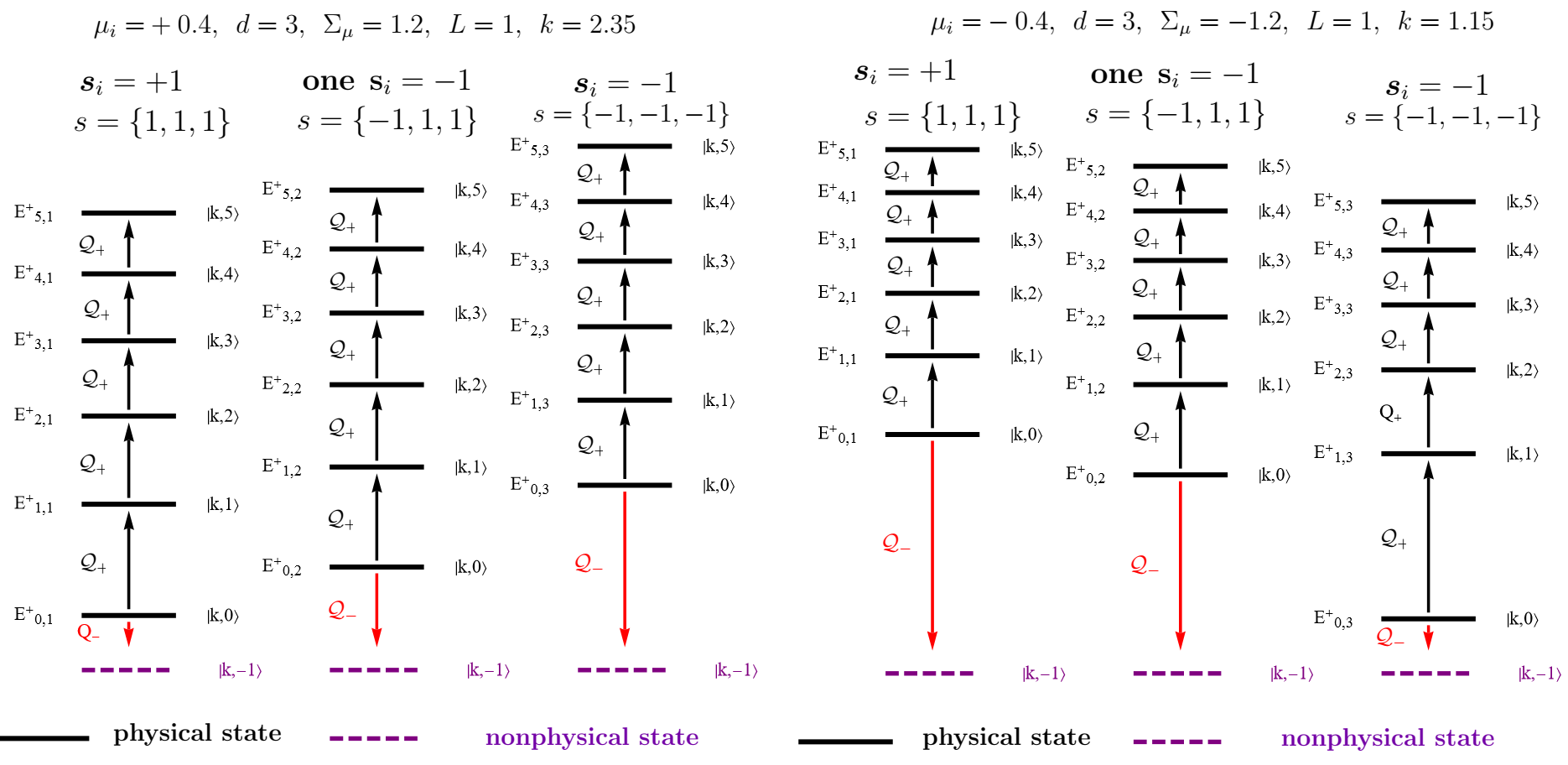}
   \caption{Algebraic $\mathfrak{su}(1,1)$ towers for the Dunkl--Klein--Gordon oscillator with $d=3$, $\mu_i=\pm0.4$, $\Sigma{\mu}=\pm1.2$, $L=1$, and $k=(2.35,1.15)$. The dashed level denotes the nonphysical state $|k,-1\rangle$. The opposite ordering of the towers reflects the spectral inversion induced by the Dunkl reflection term.} \label{fig1}
\end{figure}

Figure \ref{fig1} displays the algebraic organization of the positive-energy levels of the Dunkl--Klein--Gordon oscillator in $d=3$ dimensions within the positive discrete representation of the $\mathfrak{su}(1,1)$ algebra. Each vertical column represents an algebraic tower associated with a particular reflection sector, while the physical states $|k,n\rangle$ are generated from the lowest-weight state $|k,0\rangle$ through the repeated action of the raising operator $\mathcal{Q}_{+}$. The lowering operator satisfies the condition $\mathcal{Q}_{-}|k,0\rangle=0$, whereas the dashed level represents the formal nonphysical state $|k,-1\rangle$, included only to indicate the lower bound of the representation.

The left and right panels correspond to the isotropic Dunkl deformations $\mu_i=+0.4$ and $\mu_i=-0.4$, respectively, for the reflection sectors $s=(+1,+1,+1)$, $(-1,+1,+1)$, and $(-1,-1,-1)$. Each tower starts at a different ground-state energy due to the contribution of the Dunkl reflection term $-\sum_i \mu_i s_i$. For $\mu_i>0$, the towers shift toward higher energies as the number of negative reflections increases, whereas for $\mu_i<0$ the opposite behavior is observed, producing a spectral inversion between the different sectors. The progressive compression of the upper levels reflects the nonlinear relativistic dependence of the spectrum on the radial quantum number $n$.

The radial function \(F(\rho)\) associated with Eq.~(\ref{EDSOPR}) is obtained from the following differential equation
\begin{equation}\label{eqdsol1}
\left[
\frac{d^2}{dx^2}
+4n+2\alpha+2-x^2
+\frac{1/4-\alpha^2}{x^2}
\right]y_{n}(x)
=0,
\end{equation}
which has the solution~\cite{Lebedev1972, GurMann2005}
\begin{equation}\label{soldi2}
y_{n}(x)
=
N_{n}
e^{-x^2/2}
x^{\alpha+1/2}
L_{n}^{\alpha}(x^2).
\end{equation}
Here, $n=0,1,2,\ldots$ denotes the radial quantum number, $L_{n}^{\alpha}(x^2)$ is the generalized Laguerre polynomial of degree $n$ and parameter $\alpha=2L+\Sigma_{\mu}+d/2-1$, and $N_n$ denotes the normalization coefficient. Using the normalization condition $\int_{0}^{\infty}|y_n(x)|^2dx=1$, this coefficient is written as
\begin{equation}\label{NOR}
N_n
=
\left[
\frac{2n!}{\Gamma(n+\alpha+1)}
\right]^{1/2}.
\end{equation}
Using the solution in Eqs.~(\ref{soldi2}) and (\ref{NOR}), together with the transformation in Eq.~(\ref{TRAF}) and the dimensionless variable defined in Eq.~(\ref{DIM}), the normalized
physical radial functions can be written as
\begin{equation}\label{FUNONS}
R_{n,L}(r)
=
\left[
\frac{
2(m\omega)^{2L+\Sigma_{\mu}+d/2}n!
}{
\Gamma\!\left(n+2L+\Sigma_{\mu}+d/2\right)
}
\right]^{1/2}
r^{2L}e^{-m\omega r^2/2}
L_n^{\,2L+\Sigma_{\mu}+d/2-1}(m\omega r^2).
\end{equation}

\begin{figure}[H]
    \centering
    \includegraphics[width=0.98\textwidth]{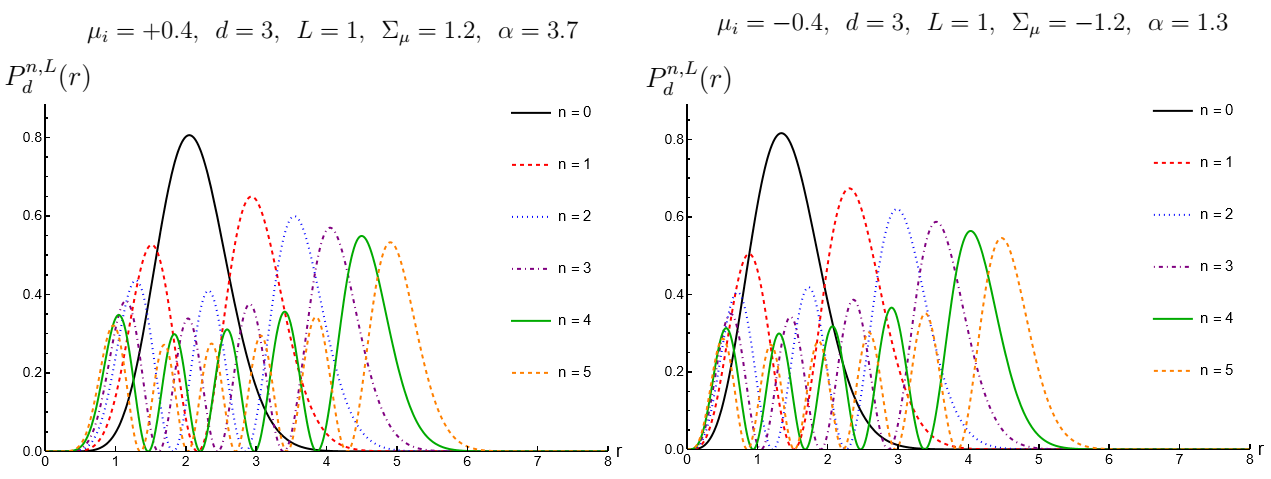}
   \caption{Physical radial probability density $P_d^{n,L}(r)=r^{d-1+2\Sigma_{\mu}}|R_{n,L}(r)|^2$ of the Sturmian basis for $\mu_i=\pm0.4$,
$d=3$, $L=1$, $\Sigma_{\mu}=\pm1.2$, and $\alpha=(3.7,1.3)$, with $n=0,\ldots,5$.} \label{fig2}
\end{figure}
Figure~\ref{fig2} displays the physical radial probability densities $P_d^{n,L}(r)=r^{d-1+2\Sigma_{\mu}}|R_{n,L}(r)|^2$ for $d=3$, $L=1$, $n=0,\ldots,5$, and $\mu_i=\pm0.4$. As the radial quantum number $n$
increases, the dominant maximum shifts toward larger values of $r$, while the number of nodes increases accordingly.

For $\Sigma_\mu>0$ $(\mu_i=+0.4)$, the probability distributions are more radially extended and their maxima occur farther from the origin. In contrast, for $\Sigma_\mu<0$ $(\mu_i=-0.4)$, the densities remain more
localized and exhibit a narrower radial support. The characteristic nodal structure of the excited states is preserved in both deformation sectors.

The corresponding calculations for $d=5$ and $d=7$ exhibit the same qualitative behavior, with increasing dimensionality producing an additional outward displacement and broadening of the distributions. Since these cases
do not introduce new qualitative features, their plots have been omitted to avoid repetition and reduce the overall length of the manuscript.

The deformation-dependent radial displacement can be quantified through the mean-square radius. Using the normalized Sturmian function in Eq.~(\ref{FUNONS}) together with the Dunkl radial measure, this quantity can be written as
\begin{equation}
\left\langle r^2\right\rangle_{n,L}
=
\frac{
2n+2L+\Sigma_{\mu}+d/2
}{
m\omega
}.
\label{oscillator-mean-square-radius}
\end{equation}
For the parameters considered here, namely $m=\omega=1$, $L=1$, and $n=0$, the root-mean-square radius for $\mu_i=+0.4$ increases from $\sqrt{4.7}\simeq2.17$ at $d=3$ to $\sqrt{8.3}\simeq2.88$ at $d=7$. For $\mu_i=-0.4$, it changes from $\sqrt{2.3}\simeq1.52$ to $\sqrt{2.7}\simeq1.64$. These values quantify the stronger outward displacement produced by positive deformation and confirm the localization trends displayed in Figure~\ref{fig2}.

\section{$\mathrm{SU}(1,1)$ radial coherent states in $d$ dimensions}
The $\mathrm{SU}(1,1)$ Perelomov coherent states are generated through the action of the displacement operator on the lowest normalized state of the corresponding unitary irreducible representation. Explicitly, they are given by
\begin{equation}
\label{eq_coh1}
|\xi\rangle
=
D(\xi)\,|k,0\rangle
=
(1-|\xi|^{2})^{k}
\sum_{n=0}^{\infty}
\left[
\frac{\Gamma(n+2k)}
{n!\,\Gamma(2k)}
\right]^{1/2}
\xi^{n}\,|k,n\rangle .
\end{equation}
where $k$ denotes the Bargmann index associated with the $\mathfrak{su}(1,1)$ algebra. Applying the displacement operator to the radial eigenfunctions, the coherent radial wave function reads
\begin{equation}
\label{eq_coh2}
R(r,\xi)
=
\left[
\frac{
2(m\omega)^{2k}(1-|\xi|^2)^{2k}
}{
\Gamma(2k)
}
\right]^{1/2}
r^{2L}e^{-m\omega r^2/2}
\sum_{n=0}^{\infty}
\xi^n
L_n^{2k-1}(m\omega r^2).
\end{equation}
To evaluate the series, we employ the generating function of the Laguerre polynomials
\begin{equation}
\label{eq_coh3}
\sum_{n=0}^{\infty}
L^{\nu}_{n}(x)\,y^{n}
=
\frac{1}{(1-y)^{\nu+1}}
\exp\!\left(-\frac{x\,y}{1-y}\right),
\qquad |y|<1.
\end{equation}
By employing this identity, the coherent-state expression in Eq.~(\ref{eq_coh2}) can be recast into a compact closed form, namely
\begin{equation}
\label{eq_coh_final}
R(r,\xi)
=
\left[
\frac{
2(m\omega)^{2k}(1-|\xi|^2)^{2k}
}{
\Gamma(2k)(1-\xi)^{4k}
}
\right]^{1/2}
r^{2L}
\exp\left[
\frac{m\omega r^2}{2}
\left(
\frac{\xi+1}{\xi-1}
\right)
\right],
\qquad |\xi|<1.
\end{equation}
Finally, by expressing the result in terms of the generalized angular momentum quantum number \(L\), the dimensionality \(d\), and the Wigner deformation parameters \(\mu_{1},\mu_{2},\ldots,\mu_{d}\), the $\mathrm{SU}(1,1)$ radial coherent states for the $d$-dimensional Dunkl--Klein--Gordon oscillator can be written as
\begin{align}
R(r,\xi)
={}&
\left[
\frac{
2(1-|\xi|^2)^{\,2L+\Sigma_\mu+d/2}
}{
\Gamma\left(2L+\Sigma_\mu+d/2\right)
(1-\xi)^{\,4L+2\Sigma_\mu+d}
}
\right]^{1/2}
\nonumber\\
&\times
(m\omega)^{\,L+\Sigma_\mu/2+d/4}
r^{2L}
\exp\left[
\frac{m\omega r^2}{2}
\left(
\frac{\xi+1}{\xi-1}
\right)
\right],
\qquad |\xi|<1.
\label{oscillator-coherent-L}
\end{align}
The coherent states preserve the standard radial $\mathrm{SU}(1,1)$ structure, while the effects of dimensionality and Dunkl deformation enter through the Bargmann index and the corresponding
weighted radial measure.
\begin{figure}[H]
    \centering
    \includegraphics[width=1.00\textwidth]{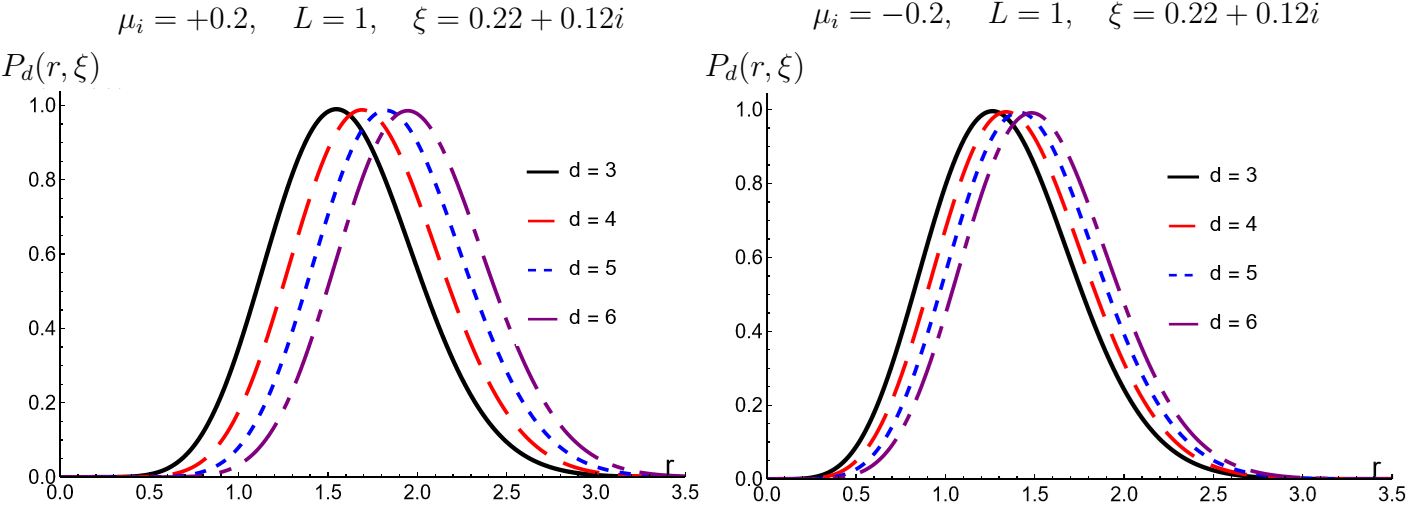}
    \caption{Physical radial coherent-state probability density
    $P_d(r,\xi)=r^{d-1+2\Sigma_{\mu}}|R(r,\xi)|^2$ for
    $\mu_i=\pm0.2$, $L=1$, $\xi=0.22+0.12i$, and $d=3,\ldots,6$.}
    \label{fig3}
\end{figure}
Figure~\ref{fig3} displays the physical radial coherent-state probability densities, defined by $P_d(r,\xi)=r^{d-1+2\Sigma_\mu}|R(r,\xi)|^2$. The results correspond to $\mu_i=\pm0.2$, $L=1$, $\xi=0.22+0.12i$, and $d=3,\ldots,6$. As the dimension increases, the probability maximum shifts toward larger radial distances, while the coherent profile becomes
progressively broader.

For positive deformation parameters $(\mu_i>0)$, the coherent-state distributions are more radially extended, and their maxima move farther from the origin as $d$ increases. In contrast, for negative deformation parameters $(\mu_i<0)$, the densities remain comparatively localized near the origin and preserve a more compact radial profile, although a moderate outward displacement is still observed with increasing dimensionality.

The calculations performed for $\mu_i=\pm0.3$ and $\pm0.4$ exhibit the same qualitative behavior. Increasing the magnitude of a positive deformation enhances the outward displacement and broadening of the coherent packet,
whereas increasing the magnitude of a negative deformation reinforces its localization near the origin.

\section{Time evolution of $\mathrm{SU}(1,1)$ radial coherent states in $d$ dimensions}
The temporal behavior of the coherent states is governed by the underlying algebraic symmetry of the model. Since the operator $K_{0}$ is related to the radial Hamiltonian through $K_{0}=\tfrac{1}{4}H_{\rho}$, the
corresponding evolution operator can be written as
\begin{equation}
U(t)
=
e^{-iH_{\rho}t/\hbar}
=
e^{-4iK_{0}t/\hbar}.
\label{Ut_d}
\end{equation}
Here, $t$ denotes the parameter governing the dynamical evolution of the system. Within this framework, the Perelomov coherent state acquires an explicit time dependence and takes the form
\begin{equation}
|\xi(t)\rangle
=
U(t)|\xi\rangle
=
U(t)D(\xi)U^{\dagger}(t)\,
U(t)|k,0\rangle.
\label{zetat_def_d}
\end{equation}
Since the lowest-weight state satisfies $K_{0}|k,0\rangle=k|k,0\rangle$, the action of the evolution operator on this state is written as
\begin{equation}
U(t)|k,0\rangle
=
e^{-4ikt/\hbar}|k,0\rangle.
\label{ground_d}
\end{equation}

Using the commutation relations
$[K_{0},\mathcal{Q}_{\pm}]=\pm\mathcal{Q}_{\pm}$ and the
Baker--Campbell--Hausdorff formula, the similarity transformations of the ladder operators take the form
\begin{equation}
U(t)\mathcal{Q}_{+}U^{\dagger}(t)
=
e^{-4it/\hbar}\mathcal{Q}_{+},
\qquad
U(t)\mathcal{Q}_{-}U^{\dagger}(t)
=
e^{4it/\hbar}\mathcal{Q}_{-}.
\label{ladder_d}
\end{equation}
Consequently, the time evolution of the displacement operator can be
written as
\begin{equation}
U(t)D(\xi)U^{\dagger}(t)
=
D\!\left(\xi(t)\right),
\qquad
\xi(t)
=
\xi e^{-4it/\hbar}.
\label{disp_d}
\end{equation}
Thus, the temporal dependence is entirely encoded in the coherent-state parameter $\xi(t)$. Since $|\xi(t)|=|\xi|$, the corresponding disentangled normal-ordered representation takes the form
\begin{equation}
D\!\left(\xi(t)\right)
=
\exp\!\left[\xi(t)\mathcal{Q}_{+}\right]
\exp\!\left[\eta K_{0}\right]
\exp\!\left[-\xi^{*}(t)\mathcal{Q}_{-}\right],
\qquad
\eta
=
\ln\!\left(1-|\xi|^{2}\right).
\label{normal_d}
\end{equation}
Combining Eqs.~\eqref{ground_d} and \eqref{normal_d}, the time-dependent Perelomov coherent state can finally be expressed as
\begin{equation}
|\xi(t)\rangle
=
e^{-4ikt/\hbar}
\exp\!\left[\xi(t)\mathcal{Q}_{+}\right]
\exp\!\left[\eta K_{0}\right]
\exp\!\left[-\xi^{*}(t)\mathcal{Q}_{-}\right]
|k,0\rangle.
\label{zetat_final_d}
\end{equation}
With the previous algebraic ingredients at hand, the radial coherent state in configuration space acquires an explicit time dependence in \(d\) dimensions and can therefore be expressed in the form
\begin{align}
\label{Rtd}
R(r,\xi(t))
={}&
e^{-4ikt/\hbar}
\left[
\frac{
2(1-|\xi|^2)^{\,2L+\Sigma_\mu+d/2}
}{
\Gamma\left(2L+\Sigma_\mu+d/2\right)
\left(1-\xi e^{-4it/\hbar}\right)^{\,4L+2\Sigma_\mu+d}
}
\right]^{1/2}
\nonumber\\
&\times
(m\omega)^{\,L+\Sigma_\mu/2+d/4}
r^{2L}
\exp\left[
\frac{m\omega r^2}{2}
\left(
\frac{
1+\xi e^{-4it/\hbar}
}{
\xi e^{-4it/\hbar}-1
}
\right)
\right],
\qquad |\xi|<1.
\end{align}
\begin{figure}[H]
    \centering
    \includegraphics[width=1.00\textwidth]{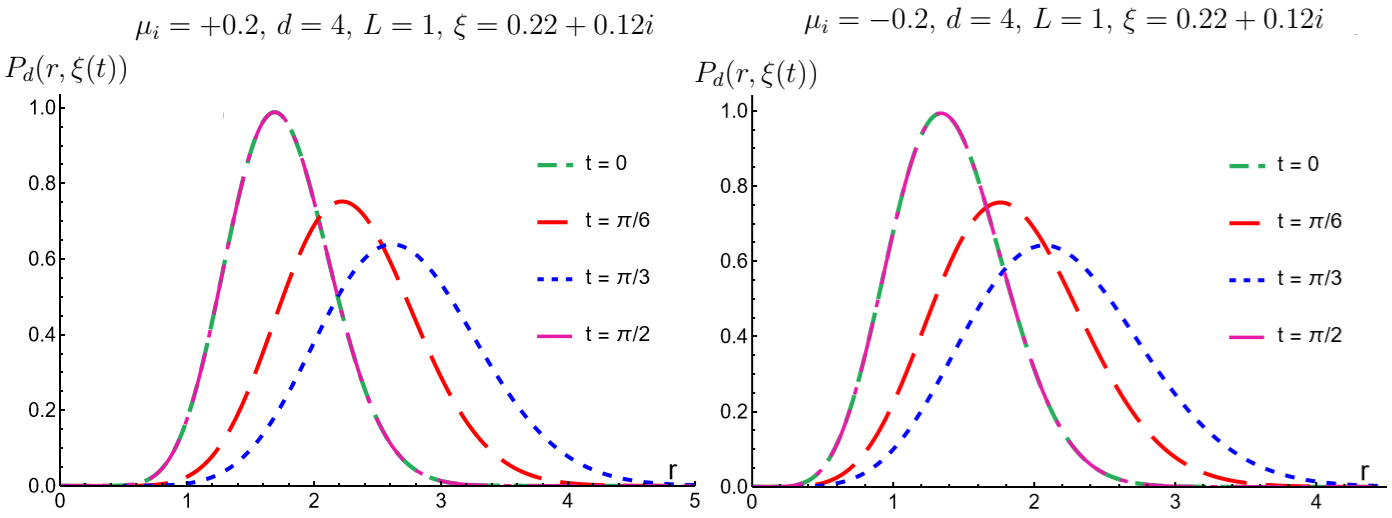}
    \caption{Time evolution of the physical radial coherent-state probability density $P_d(r,\xi(t))=r^{d-1+2\Sigma_{\mu}}|R(r,\xi(t))|^{2}$ for $\mu_i=\pm0.2$,
$d=4$, $L=1$, $\xi=0.22+0.12i$, and $t=0,\pi/6,\pi/3,\pi/2$.}
\label{fig4}
\end{figure}
Figure~\ref{fig4} displays the time evolution of the physical radial coherent-state probability density $P_d(r,\xi(t))$ for $d=4$, $L=1$, and the isotropic deformation sectors $\mu_i=\pm0.2$. Each curve corresponds to a fixed instant of time. From $t=0$ to $t=\pi/3$, the probability maximum moves toward larger radial distances, while the distribution becomes broader and its maximum decreases. At $t=\pi/2$, the initial profile is recovered, as shown by the exact overlap of the $t=0$ and $t=\pi/2$ curves. This recurrence follows directly from $\xi(t)=\xi e^{-4it/\hbar}$ and reveals the periodic radial breathing of the coherent packet.

The sign of the Dunkl deformation determines the radial region in which this dynamics occurs. For $\mu_i>0$, the packet is displaced toward larger radii and exhibits a greater radial extension, whereas for $\mu_i<0$ it remains closer to the origin and preserves a more localized profile. Calculations performed for $\mu_i=\pm0.3$ and $\pm0.4$ exhibit the same periodic behavior: increasing the positive deformation shifts the packet farther outward, while increasingly negative values enhance its localization near the origin.

\section{Coulomb potential: Bound states}
In this section, we analyze algebraically the bound-state sector of the $d$-dimensional Dunkl--Klein--Gordon equation in the presence of an attractive Coulomb interaction, employing once again the Schr\"odinger factorization method together with the $\mathfrak{su}(1,1)$ symmetry framework. In particular, the interaction is introduced through the potential
\begin{equation}
V(r) = -\frac{Ze^2}{r}.
\end{equation}
This interaction is incorporated into the radial equation through the relativistic substitution $E^2 \to \left(E + \frac{Ze^2}{r}\right)^2$, which introduces both Coulombic and inverse-square contributions. As a result, the radial equation takes the form
\begin{equation}\label{DIFSECOCP}
\left[
\frac{d^2}{dr^2}
+ \frac{d-1+2(\mu_1+\mu_2+\cdots+\mu_d)}{r}\frac{d}{dr}
+ (E^2 - m^2)
+ \frac{2EZe^2}{r}
+ \frac{Z^2 e^4 - \varpi^2}{r^2}
\right] R(r)=0.
\end{equation}
This equation describes the bound-state sector \((E^{2}<m^{2})\) and incorporates the combined effects of the dimensionality, the Dunkl deformation parameters, and the effective angular momentum contribution, thus providing the natural starting point for the algebraic construction of normalizable solutions. A closed set of $\mathfrak{su}(1,1)$ generators is constructed by adopting the ansatz
\begin{equation}\label{FACS}
\left[\rho\frac{d}{d\rho}+\mathscr{A}\rho+\mathscr{B}\right]\biggl[-\rho\frac{d}{d\rho}+\mathscr{C}\rho+\mathscr{F}\biggr]F(\rho)=\mathscr{G}F(\rho).
\end{equation}
A direct identification of Eq.~(\ref{FACS}) with Eq.~(\ref{DIFSECOCP}) results in the parameters $\mathscr{A}$, $\mathscr{B}$, $\mathscr{C}$, $\mathscr{F}$, and $\mathscr{G}$ given in Table~$2$.
\begin{table}[H]
\centering
\caption{Schr\"odinger factorization parameters for $d$-dimensional Dunkl--Klein--Gordon with a Dunkl-Coulomb-like potential.}
\label{tab:ABCFGDKG2}
\resizebox{0.95\textwidth}{!}{
\begin{tabular}{ccccc}
\hline\hline
$\mathscr{A}$ & $\mathscr{B}$ & $\mathscr{C}$ & $\mathscr{F}$ & $\mathscr{G}$ \\
\hline
$\sqrt{m^2-E^2}$ & $-\mathscr{L}_0+\frac{\alpha_d}{2}-1$ & $\sqrt{m^2-E^2}$ & $-\mathscr{L}_0+\frac{\alpha_d}{2}$
& $\mathscr{L}_0^2+\mathscr{L}_0-\frac{\alpha_d}{2}\left(\frac{\alpha_d}{2}-1\right)+Z^2e^4-\bar{\omega}^2$ \\
$-\sqrt{m^2-E^2}$ & $\mathscr{L}_0+\frac{\alpha_d}{2}-1$ & $-\sqrt{m^2-E^2}$ & $\mathscr{L}_0+\frac{\alpha_d}{2}$
& $\mathscr{L}_0^2-\mathscr{L}_0-\frac{\alpha_d}{2}\left(\frac{\alpha_d}{2}-1\right)+Z^2e^4-\bar{\omega}^2$ \\
\hline\hline
\end{tabular}
}
\end{table}
where
\begin{equation}
\mathscr{L}_0=\frac{EZe^2}{\sqrt{m^2-E^2}},\hspace{0.2cm} \alpha_d = d - 1 + 2\mu_1 + 2\mu_2 + 2\mu_3 + \cdots + 2\mu_d.
\end{equation}
These parameters are determined by expanding Eq.~(\ref{FACS}) and performing a term-by-term identification with Eq.~(\ref{DIFSECOCP}). It then follows that the equation satisfied by $R(r)$ can be cast in the factorized form
\begin{equation}
\left(\mathcal{L}_{\mp}\mp 1\right)\mathcal{L}_{\pm}=\mathscr{L}_0^2\pm\mathscr{L}_0-\frac{\alpha_d}{2}\left(\frac{\alpha_d}{2}-1\right)+Z^2e^4-\bar{\omega}^2,
\end{equation}
where
\begin{equation}
\mathcal{L}_{\pm}=\mp r\frac{d}{dr}+\sqrt{m^2-E^2}r-\frac{EZe^2}{\sqrt{m^2-E^2}}\mp\frac{\alpha_d}{2}.
\end{equation}
These results allow us to construct two new operators

\begin{align}\label{OPSB1}
\mathcal{G}_{\pm}=\mp r\frac{d}{dr}+\sqrt{m^2-E^2}r\mp\frac{\alpha_d}{2}-\mathcal{H}_0,
\end{align}
here
\begin{equation}\label{TEROPC1}
\mathcal{H}_0 R(r)=\frac{1}{2\sqrt{m^2-E^2}}
\left[
-r\frac{d^2}{dr^2}
- \alpha_d\frac{d}{dr}
- E^2r+ m^2r
+\frac {\varpi^2-Z^2 e^4 }{r}
\right] R(r) = \mathscr{L}_0R(r).
\end{equation}
Eqs.~(\ref{OPSB1}) and (\ref{TEROPC1}) are readily seen to define a closed realization of the $\mathfrak{su}(1,1)$ algebra. In order to derive the energy spectrum associated with the bound-state sector of the \(d\)-dimensional Dunkl--Klein--Gordon equation in the presence of an attractive Coulomb interaction, we employ the unitary irreducible representations of the noncompact $\mathfrak{su}(1,1)$ Lie algebra. In this framework, the quadratic Casimir operator \(\mathbb{C}^{2}\) satisfies the eigenvalue relation
\begin{equation}\label{CASC1COU}
\mathbb{C}^2=\mathcal{H}_{0}
\left(
\mathcal{H}_{0}-1
\right)
-
\mathcal{G}_{+}\mathcal{G}_{-}
=
\varpi^{2}-Z^{2}e^{4}
+
\left(
\frac{\alpha_d-1}{2}
\right)^{2}
-\frac{1}{4}=k(k-1),
\end{equation}
by combining Eqs.~(\ref{OMEGAL}) and (\ref{CASC1COU}), the Bargmann index $k^{C}$ associated with the Coulomb sector can be written as
\begin{align}\label{BARGCOU}\nonumber
k^{C} &=
\frac{1}{2}
+
\sqrt{
\varpi^2
+
\left(\sum_{i=1}^d\mu_i\right)
\left(\sum_{i=1}^d\mu_i+d-2\right)
+
\left(\frac d2-1\right)^2
-
Z^2e^4
}\\
&=
\frac{1}{2}
+
\sqrt{
\left(
2L+\sum_{i=1}^d\mu_i+\frac{d-2}{2}
\right)^2
-
Z^{2}e^{4}
}.
\end{align}
The precise condition for the Bargmann index to be real is therefore
\begin{equation}\label{REALCOND}
\left|
2L+\sum_{i=1}^d\mu_i+\frac{d-2}{2}
\right|
\geq |Ze^2|.
\end{equation}
The strict inequality corresponds to $k_L^{C}>1/2$, whereas equality gives the limiting value $k_L^{C}=1/2$. Both cases yield a real Bargmann index compatible with the positive-discrete representation considered here, while equality defines the critical boundary beyond which the Bargmann index becomes complex. The additional normalizability and bound-state conditions are discussed in Sec.~7.

The dependence on the Dunkl deformation is governed by the signed sum $\Sigma_{\mu}=\sum_{i=1}^{d}\mu_i$, rather than by the individual magnitudes $|\mu_i|$. Since this quantity enters through the quadratic combination
\begin{equation}
\left(
2L+\Sigma_{\mu}+\frac{d-2}{2}
\right)^2,
\end{equation}
the reality condition is not, in general, monotonic with respect to the magnitude of the deformation. This behavior becomes particularly transparent in the isotropic case $\mu_i=\mu$, for which $\Sigma_{\mu}=d\mu$. For positive deformation, $\mu=|\mu|$, the quantity $2L+d|\mu|+\frac{d-2}{2}$ increases with $|\mu|$, and therefore its square increases monotonically.

For negative deformation, $\mu=-|\mu|$, however, the corresponding quantity becomes $2L-d|\mu|+\frac{d-2}{2}$. Consequently, its square may initially decrease as $|\mu|$ increases, reach a minimum, and subsequently increase again.

Thus, the reality of the Coulomb Bargmann index cannot, in general, be inferred from $|\mu|$ alone, but must be determined from the signed combination $2L+\Sigma_{\mu}+(d-2)/2$ for each sector specified by $d$, $L$, $\Sigma_{\mu}$, and the Coulomb coupling $Ze^2$. Combining Eqs.~(\ref{TEROPC1}), (\ref{OPTLA}) and (\ref{BARGCOU}) leads straightforwardly to the quantization condition
\begin{equation}
\frac{E^{C}Ze^2}{\sqrt{m^2-\left(E^{C}\right)^2}}
=
n+\frac{1}{2}+\sqrt{
\left(
2L+\sum_{i=1}^d\mu_i+\frac{d-2}{2}
\right)^2
-
Z^{2}e^{4}
},
\end{equation}
from which the energy spectrum associated with the \(d\)-dimensional Dunkl--Klein--Gordon equation in the presence of an attractive Coulomb interaction is derived
\begin{equation}
E_{n,L}^{C}
=
\pm m
\left[
1+
\frac{Z^2e^4}{
\left[
n+\frac{1}{2}
+
\sqrt{
\left(
2L+\Sigma_\mu+\frac{d-2}{2}
\right)^2-Z^2e^4
}
\right]^2
}
\right]^{-1/2}.
\label{energy-coulomb}
\end{equation}

\begin{figure}[H]
    \centering
    \includegraphics[width=0.98\textwidth]{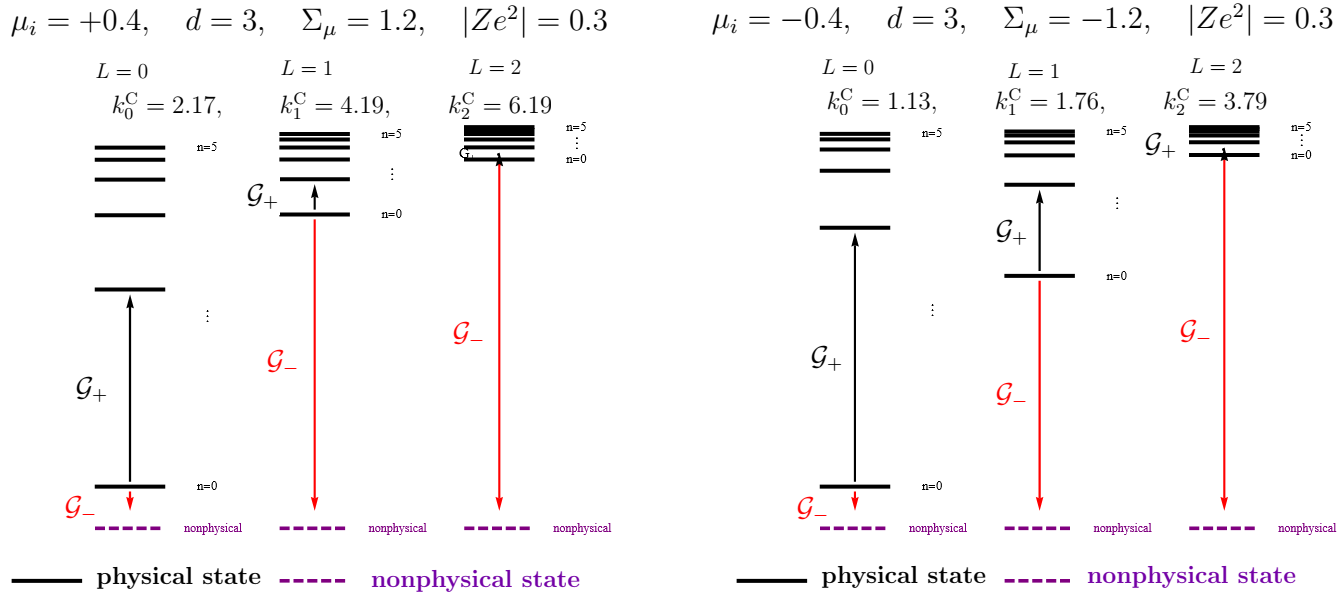}
   \caption{
Algebraic tower structure of the positive-energy spectrum of the $d=3$ Dunkl--Klein--Gordon Coulomb problem for the isotropic deformation
parameters $\mu_i=\pm0.4$ and Coulomb coupling $|Ze^2|=0.3$. For the parameters considered, the sectors $L=0,1,2$ satisfy the reality condition
for the Bargmann index and generate physical bound-state towers within the positive discrete representation of $\mathfrak{su}(1,1)$. The dashed level
below each tower denotes the formal nonphysical state $|k_L,-1\rangle$.}\label{fig5}
\end{figure}

Figure~\ref{fig5} displays the algebraic organization of the positive-energy spectrum of the $d=3$ Dunkl--Klein--Gordon Coulomb problem within the positive discrete representation of the $\mathfrak{su}(1,1)$ algebra. The left and right panels correspond to the isotropic deformation parameters $\mu_i=+0.4$ and $\mu_i=-0.4$, respectively, with $|Ze^2|=0.3$. Each vertical column represents an algebraic tower associated with a fixed generalized
angular momentum value $L=0,1,2$. For the parameters considered, all the displayed sectors satisfy the reality condition given in Eq.~(\ref{REALCOND}), and hence possess real Bargmann indices and define physical positive-discrete representations.

The physical states are generated from the lowest-weight state $|k_L,0\rangle$ through repeated application of the raising operator $\mathcal{G}_{+}$, while the lowering operator satisfies $\mathcal{G}_{-}|k_L,0\rangle=0$. The dashed purple level below each tower represents the formal state $|k_L,-1\rangle$, which lies outside the physical representation and is included only to indicate its algebraic lower bound.

Unlike the oscillator spectrum, the Coulomb spectrum does not depend explicitly on the reflection eigenvalues $s_i$. The towers are therefore classified by $L$, whereas the effect of the Dunkl deformation is encoded in $\Sigma_\mu$. Increasing $L$ modifies the effective angular contribution and the Bargmann index $k_L$, thereby changing the position and spacing of the levels within each tower.

The nonuniform separation between consecutive levels reflects the relativistic character of the Coulomb spectrum. The low-lying states are more widely separated, whereas the highly excited states become progressively compressed and accumulate near the positive-energy continuum threshold $E\to+m$. The figure therefore illustrates both the ladder structure generated by $\mathcal{G}_{\pm}$ and the combined effects of the generalized angular momentum and Dunkl deformation on the bound-state spectrum.

The explicit form of the radial wave function \(R(r)\) follows directly from the standard second-order differential equation framework \cite{Lebedev1972}
\begin{equation}\label{ECDCOU1}
x\,u'' + (\sigma + 1 - 2\nu)\,u' +
\left[
n + \frac{\sigma + 1}{2}
- \frac{x}{4}
+ \frac{\nu(\nu - \sigma)}{x}
\right] u = 0,
\end{equation}
which has the particular solution
\begin{equation}
u(x)=N_n\,e^{-x/2}\,x^{\nu}\,L_n^{\sigma}(x),
\end{equation}
Introducing the transformation \(x = 2ar\) into Eq.~(\ref{DIFSECOCP}) and matching the resulting equation with Eq.~(\ref{ECDCOU1}) leads to the following identifications
\begin{equation}
\sigma + 1 - 2\nu = d-1 + 2\sum_{i=1}^d \mu_i, \quad n + \frac{\sigma + 1}{2}= \frac{E Z e^2}{a}, \quad \nu(\nu - \sigma) = Z^2 e^4 - \bar{\omega}^2.
\end{equation}
As a result, the radial sector is described by the following explicit wave functions \(R(r)\),
\begin{equation}\label{FUNONBC}
R_n(r)=
\sqrt{
\frac{(2a)^{\,d+2\Sigma_\mu}\, n!}
{\left(2n+\sigma+1\right)\,\Gamma(n+\sigma+1)}
}
\, e^{-a r}
(2a r)^{\nu}
L_n^{\sigma}(2a r),
\end{equation}
where
\begin{equation}\label{NUCOU}
\nu=
\frac{
-\left(d-2+2\displaystyle\sum_{i=1}^d \mu_i\right)
+\sqrt{\left(d-2+2\displaystyle\sum_{i=1}^d \mu_i\right)^2
+4\left(\bar{\omega}^2-Z^2e^4\right)}
}{2},
\end{equation}
and
\begin{equation}\label{SIGCOU}
\sigma=
\sqrt{
\left(d-2+2\sum_{i=1}^d \mu_i\right)^2
+4\left(\bar{\omega}^2-Z^2e^4\right)
},
\end{equation}
for bound states satisfying $E^2<m^2$, we introduce the positive decay parameter $a=\sqrt{m^2-E^2}$.
\begin{figure}[H]
    \centering
    \includegraphics[width=0.98\textwidth]{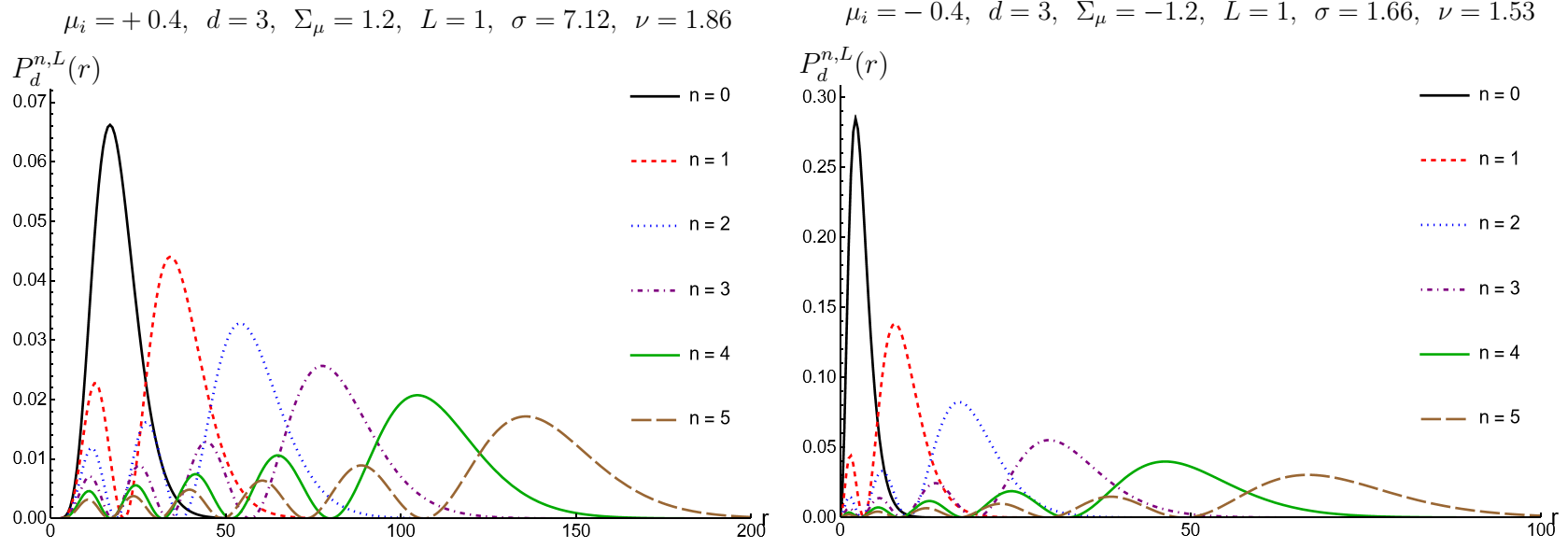}
   \caption{Physical radial probability density $P_d^{n,L}(r)=r^{d-1+2\Sigma_\mu} |R_{n,L}^{\mathrm C}(r)|^2$ of the Sturmian Coulomb basis for $\mu_i=\pm0.4$, $d=3$,
$\Sigma_{\mu}=\pm1.2$, $L=1$, $\sigma=(7.12,1.66)$ and $\nu=(1.86,1.53) $ with $n=0,\ldots,5$.} \label{fig6}
\end{figure}
Figure~\ref{fig6} displays the physical radial probability densities $P_d^{n,L}(r)=r^{d-1+2\Sigma_\mu} |R_{n,L}^{\mathrm C}(r)|^2$ of the Coulomb Sturmian basis for $d=3$, $L=1$, $n=0,\ldots,5$, and
$\mu_i=\pm0.4$. As the radial quantum number $n$ increases, the dominant probability maximum shifts toward larger radial distances and additional nodes appear, reflecting the characteristic Sturmian structure of the
Coulomb states.

For $\Sigma_\mu>0$ $(\mu_i=+0.4)$, the radial distributions are more extended and their maxima move progressively farther from the origin as $n$ increases. This spreading is more pronounced than in the oscillator
sector because the Coulomb states exhibit longer radial tails. In contrast, for $\Sigma_\mu<0$ $(\mu_i=-0.4)$, the densities remain comparatively localized near the origin and preserve a more compact radial profile,
although their maxima also shift outward with increasing $n$.

The corresponding calculations for $d=5$ and $d=7$ exhibit the same qualitative behavior. Increasing the dimension enhances the outward displacement and radial spreading for $\Sigma_\mu>0$, whereas the
distributions remain comparatively localized for $\Sigma_\mu<0$.

For the $d$-dimensional Dunkl--Klein--Gordon equation with an attractive Coulomb interaction, the $\mathrm{SU}(1,1)$ Perelomov coherent states are generated by applying the displacement operator to the normalized ground state of the representation space, and are given by
\begin{equation}
R(r,\xi)
=
\left[
\frac{
(2a)^{d+2\Sigma_\mu}
\left(1-|\xi|^2\right)^{\sigma+2}
}{
\Gamma(\sigma+2)
|1-\xi|^{2\sigma+4}
}
\right]^{1/2}
(2ar)^\nu
\exp\left[
-ar\left(\frac{1+\xi}{1-\xi}\right)
\right],
\qquad |\xi|<1.
\label{coulomb-physical-coherent-state}
\end{equation}
with the parameters \(k^{C}\), \(\nu\), and \(\sigma\) determined by Eqs.~(\ref{BARGCOU}), (\ref{NUCOU}), and (\ref{SIGCOU}), respectively.
\begin{figure}[H]
    \centering
    \includegraphics[width=0.98\textwidth]{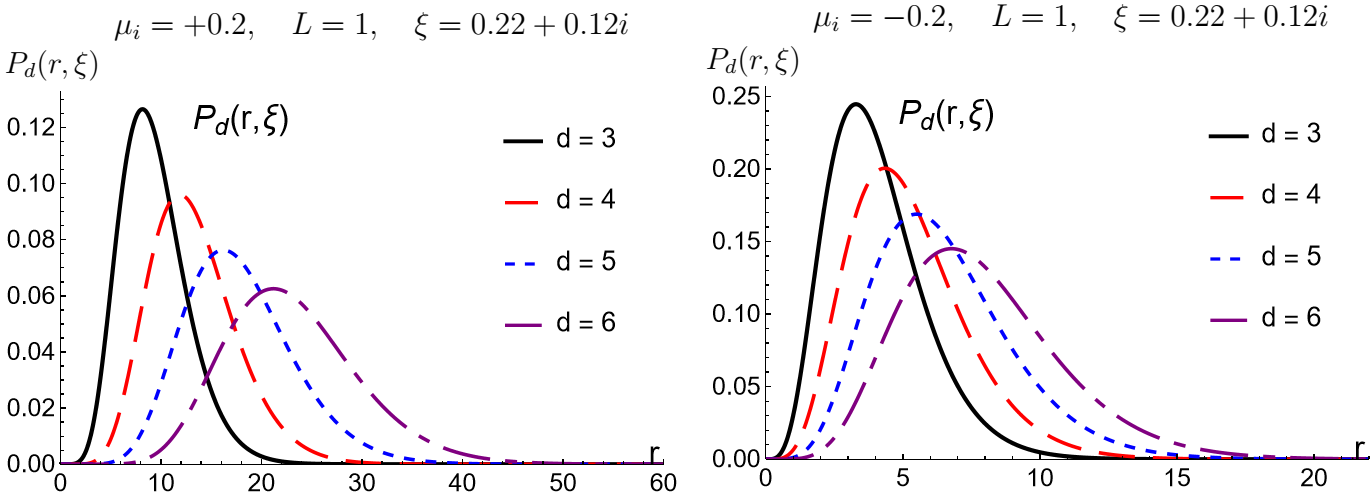}
  \caption{Physical radial coherent-state probability density $P_d(r,\xi)=r^{d-1+2\Sigma_{\mu}}|R(r,\xi)|^2$ for $\mu_i=\pm0.2$, $L=1$
$\xi=0.22+0.12i$, and $d=3,\ldots,6$.} \label{fig7}
\end{figure}
Figure~\ref{fig7} displays the physical radial coherent-state probability densities associated with the Coulomb interaction. The densities are defined by $P_d(r,\xi)=r^{d-1+2\Sigma_\mu}|R(r,\xi)|^2$ and are shown for
$\mu_i=\pm0.2$, $L=1$, $\xi=0.22+0.12i$, and $d=3,\ldots,6$. For either sign of the deformation, increasing the dimension shifts the probability maximum toward larger radial distances and broadens the coherent profile.

The calculations performed for $\mu_i=\pm0.3$ and $\pm0.4$, not displayed to avoid repetition, confirm that the magnitude and sign of the deformation control the radial scale of the coherent states. Increasing a positive
deformation produces a progressively stronger outward displacement and a broader radial distribution. Conversely, increasing the magnitude of a negative deformation shifts the densities toward the origin and produces
more compact profiles. The different radial and vertical scales used in the two panels reflect this pronounced difference in localization and allow each set of normalized densities to be displayed clearly.

Compared with the oscillator coherent states shown in Figure~\ref{fig3}, the Coulomb profiles exhibit a larger radial extension. The oscillator densities remain comparatively compact because of their
Gaussian decay, whereas the Coulomb coherent states possess exponential radial tails. The outward displacement and broadening are therefore more pronounced in the Coulomb sector, particularly for positive deformation
parameters and higher dimensions.
\begin{equation}
R(r,\xi(t))
=
e^{-2ik^{\mathrm C}t/\hbar}
\left[
\frac{
(2a)^{d+2\Sigma_\mu}
\left(1-|\xi(t)|^2\right)^{\sigma+2}
}{
\Gamma(\sigma+2)
|1-\xi(t)|^{2\sigma+4}
}
\right]^{1/2}
(2ar)^\nu
\exp\left[
-ar\left(
\frac{1+\xi(t)}{1-\xi(t)}
\right)
\right],
\label{coulomb-time-dependent-coherent-state}
\end{equation}
where $\xi(t)=\xi e^{-2it/\hbar}$, \(\nu\), and \(\sigma\) are defined in Eqs.~(\ref{NUCOU}) and (\ref{SIGCOU}), respectively.

\begin{figure}[H]
    \centering
    \includegraphics[width=0.98\textwidth]{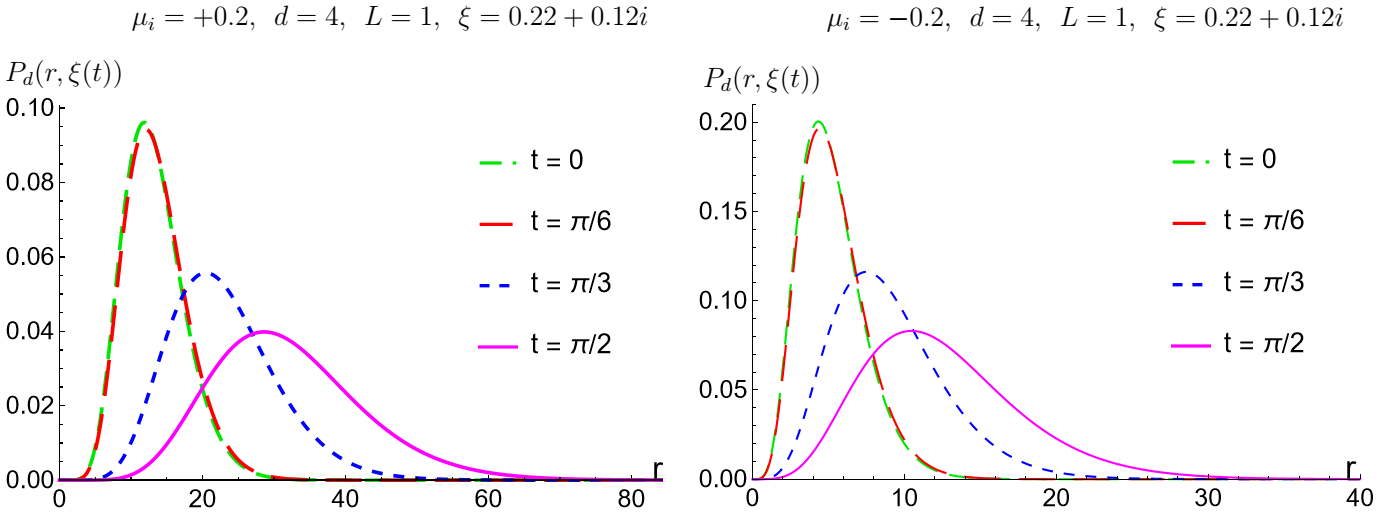}
  \caption{Time evolution of the physical radial coherent-state probability density $P_d(r,\xi(t))=r^{d-1+2\Sigma_{\mu}}|R(r,\xi(t))|^{2}$ for $\mu_i=\pm0.2$,
$d=4$, $L=1$, $\xi=0.22+0.12i$, and $t=0,\pi/6,\pi/3,\pi/2$.} \label{fig8}
\end{figure}
Figure~\ref{fig8} displays the algebraic time evolution of the physical radial coherent-state probability density $P_d(r,\xi(t))=r^{d-1+2\Sigma_\mu}|R(r,\xi(t))|^2$ for the Coulomb interaction, with $d=4$, $L=1$, $\xi=0.22+0.12i$, and $\mu_i=\pm0.2$. The curves at $t=0$ and $t=\pi/6$ nearly overlap, whereas at later times the probability maximum moves toward larger radial distances and the profile becomes broader and lower. This behavior
represents the expanding part of the periodic radial breathing generated by the phase rotation $\xi(t)=\xi e^{-2it/\hbar}$.

The calculations for $\mu_i=\pm0.3$ and $\pm0.4$, not displayed to avoid repetition, exhibit the same qualitative evolution. Increasing the positive deformation shifts the dynamics toward progressively larger radial regions
and enhances the spatial extension of the packet. Conversely, increasing the magnitude of a negative deformation keeps the evolving density closer to the origin and produces a more compact profile. The panels use different
radial and vertical scales to display clearly the corresponding normalized densities. Compared with the oscillator sector, the Coulomb evolution develops longer radial tails and a larger spatial extension because of its
exponential, rather than Gaussian, radial decay.

\section{Physical framework and consistency checks}
This section addresses the physical foundations and the main consistency checks of the present Dunkl--Klein--Gordon formulation. We first establish the weighted integration measure, the normalization conditions, and the regularity requirements of the radial solutions. We then examine the gauge, Lorentz, and reflection symmetries of the model. Finally, as consistency checks of the algebraic construction, we derive the nonrelativistic limit
of the Coulomb sector and examine the non-deformed and three-dimensional limits, comparing the resulting equations, spectra, and radial solutions with established Klein--Gordon and Dunkl--Klein--Gordon results.

\subsection{Dunkl measure and normalization}
The natural scalar product of the Dunkl formulation is defined in the
weighted Hilbert space associated with the measure
\begin{equation}
d\mu_D(\mathbf{x})
=
w_\mu(\mathbf{x})\,d^dx,
\qquad
w_\mu(\mathbf{x})
=
\prod_{i=1}^{d}|x_i|^{2\mu_i},
\qquad
\mu_i>-\frac{1}{2}.
\label{dunkl-cartesian-measure}
\end{equation}
The reflection operators do not enter the integration measure as multiplicative factors. Their contribution is instead encoded through the reflection-invariant weight and through the decomposition of the Hilbert space into sectors of definite reflection parity. In particular, the weight and the reflection operators satisfy
\begin{equation}
w_\mu(R_i\mathbf{x})
=
w_\mu(\mathbf{x}),
\qquad
R_i^\dagger
=
R_i,
\qquad
R_i^2
=
1.
\label{reflection-invariance-of-measure}
\end{equation}
Accordingly, the reflection operators are self-adjoint with respect to the
weighted scalar product, as expressed by
\begin{equation}
\int_{\mathbb{R}^{d}}
f^{*}(\mathbf{x})R_i g(\mathbf{x})\,d\mu_D(\mathbf{x})
=
\int_{\mathbb{R}^{d}}
\left[R_i f(\mathbf{x})\right]^{*}
g(\mathbf{x})\,d\mu_D(\mathbf{x}).
\label{reflection-self-adjointness}
\end{equation}
Since the operators $R_i$ are mutually commuting, self-adjoint involutions,
the Hilbert space decomposes into mutually orthogonal sectors of definite
reflection parity,
\begin{equation}
R_i\Psi_{\mathbf{s}}
=
s_i\Psi_{\mathbf{s}},
\qquad
s_i=\pm1.
\label{reflection-parity-sectors}
\end{equation}
Under appropriate boundary conditions, the Dunkl derivatives are
anti-symmetric and the corresponding Dunkl momentum operators are symmetric
with respect to the same weighted scalar product~\cite{Dunkl1989,Rosler2003}.

In the hyperspherical coordinates introduced in Sec.~2, the weighted
measure can be factorized as
\begin{equation}
d\mu_D(\mathbf{x})
=
r^{d-1+2\Sigma_\mu}\,dr\,d\Omega_\mu,
\label{dunkl-hyperspherical-measure}
\end{equation}
where the angular contribution takes the form
\begin{align}
d\Omega_\mu
={}&
|\cos\theta_1|^{2\mu_1}
|\sin\theta_1|^{2\mu_2}\,d\theta_1
\nonumber\\
&\times
\prod_{j=2}^{d-1}
(\sin\theta_j)^{
j-1+2\sum_{i=1}^{j}\mu_i
}
|\cos\theta_j|^{2\mu_{j+1}}\,d\theta_j.
\label{dunkl-angular-measure}
\end{align}
The generalized orthonormality condition for the complete wavefunctions
is consequently written as
\begin{equation}
\int_{\mathbb{R}^d}
\Psi_{\alpha}^{*}(\mathbf{x})
\Psi_{\alpha'}(\mathbf{x})\,
d\mu_D(\mathbf{x})
=
\delta_{\alpha\alpha'}.
\label{dunkl-complete-normalization}
\end{equation}

For the separated solutions employed in the algebraic construction, the angular functions can be chosen to satisfy
\begin{equation}
\int
\Theta_{\boldsymbol{\ell},\mathbf{s}}^{*}(\Omega)
\Theta_{\boldsymbol{\ell}',\mathbf{s}'}(\Omega)\,
d\Omega_\mu
=
\delta_{\boldsymbol{\ell}\boldsymbol{\ell}'}
\delta_{\mathbf{s}\mathbf{s}'}.
\label{dunkl-angular-orthonormality}
\end{equation}
The associated radial normalization condition then takes the form
\begin{equation}
\int_{0}^{\infty}
r^{d-1+2\Sigma_\mu}
R_{n,L}^{*}(r)R_{n',L}(r)\,dr
=
\delta_{nn'}.
\label{dunkl-radial-normalization}
\end{equation}
Accordingly, the weighted radial density used throughout the manuscript
can be written as
\begin{equation}
P_{n,L}^{(d)}(r)
=
r^{d-1+2\Sigma_\mu}|R_{n,L}(r)|^2,
\qquad
\int_{0}^{\infty}
P_{n,L}^{(d)}(r)\,dr
=
1.
\label{dunkl-radial-density-normalization}
\end{equation}
Equation~\eqref{dunkl-radial-normalization} defines the weighted radial normalization adopted for the Sturmian basis. For a fixed angular sector and a fixed radial scale, the basis functions can be chosen to satisfy this orthonormality relation. This weighted $L^2$ normalization should be distinguished from the conserved Klein--Gordon charge normalization of physical stationary solutions. For two time-dependent solutions, the corresponding Klein--Gordon inner product is written as
\begin{equation}
\left(\Psi_1,\Psi_2\right)_{\mathrm{KG}}
=
i
\int_{\mathbb{R}^d}
w_\mu(\mathbf{x})
\left[
\Psi_1^*\mathcal{D}_t\Psi_2
-
\left(\mathcal{D}_t\Psi_1\right)^*\Psi_2
\right]d^dx,
\label{dunkl-klein-gordon-inner-product}
\end{equation}
where $\mathcal{D}_t$ denotes the temporal covariant derivative in the presence of a time-like vector potential. For stationary solutions belonging to the positive-charge branch, the
overall normalization coefficient can be chosen so that the conserved Klein--Gordon charge is equal to unity. The normalization coefficient fixes the magnitude of the charge but does not alter its sign ~\cite{Greiner2000,Ryder1996}.

\subsection{Regularity conditions}
With the weighted radial density defined in
Eq.~\eqref{dunkl-radial-density-normalization}, regularity at the origin
must be examined through $P_{n,L}^{(d)}(r)$ rather than through
$|R_{n,L}(r)|^2$ alone. For the oscillator sector,
Eq.~\eqref{FUNONS} gives $R_{n,L}^{\mathrm{osc}}(r)\sim r^{2L}$ near the origin. Accordingly, the physical radial density behaves as
\begin{equation}
P_d^{\mathrm{osc}}(r)
\underset{r\rightarrow0}{\sim}
r^{\,4L+2\Sigma_\mu+d-1}.
\label{oscillator-origin-behavior}
\end{equation}
Local square integrability requires
\begin{equation}
2L+\Sigma_\mu+\frac{d}{2}>0.
\label{oscillator-regularity-condition}
\end{equation}
This condition is automatically satisfied for $L\geq0$ and
$\mu_i>-1/2$. At large distances, convergence is ensured by the Gaussian
factor $e^{-m\omega r^2/2}$.

For the Coulomb sector, Eqs.~(\ref{BARGCOU}), (\ref{FUNONBC}), and (\ref{NUCOU}) give the
near-origin behavior
\begin{equation}
R_{n,L}^{\mathrm C}(r)
\underset{r\rightarrow0}{\sim}
r^\nu,
\qquad
\nu
=
k_L^{\mathrm C}
-\Sigma_\mu
-\frac{d-1}{2},
\label{coulomb-wavefunction-origin}
\end{equation}
where $k_L^{\mathrm C}$ is the Coulomb Bargmann index. Combining this
behavior with the weighted radial measure gives
\begin{equation}
P_d^{\mathrm C}(r)
\underset{r\rightarrow0}{\sim}
r^{\,d-1+2\Sigma_\mu+2\nu}
=
r^{\,2k_L^{\mathrm C}}.
\label{coulomb-origin-behavior}
\end{equation}
Since the reality condition implies $k_L^{\mathrm C}\geq1/2$, the physical radial density vanishes at the origin and is locally integrable. Although the radial function itself may exhibit a weak singularity when
$\nu<0$, this behavior does not exclude the state, since physical admissibility is determined by square integrability with respect to the Dunkl measure.

At large distances, the factor $e^{-ar}$, with $a=\sqrt{m^2-E^2}$, ensures convergence when $|E|<m$. In addition, the quantization condition requires $EZe^2>0$ for the attractive bound-state
channel. The admissible parameter domain can therefore be summarized as
\begin{equation}
\mu_i>-\frac{1}{2},
\qquad
\left|
2L+\Sigma_\mu+\frac{d-2}{2}
\right|
\geq |Ze^2|,
\qquad
|E|<m,
\qquad
EZe^2>0.
\label{physical-admissibility-domain}
\end{equation}
The last condition selects the energy branch corresponding to the chosen
sign of the Coulomb coupling.

\subsection{Gauge and Lorentz symmetries}
The reflection structure of the Dunkl derivatives modifies the action of continuous space-time symmetries on the solution fields. Under a local gauge transformation $\Psi'(x)=g(x)\Psi(x)$, with
$g(x)=e^{iq\chi(x)}$, the failure of the Dunkl derivative to commute with the local phase transformation can be written as
\begin{equation}
D_i(g\Psi)-gD_i\Psi
=
[D_i,g]\Psi
=
\left(\partial_i g\right)\Psi
+
\frac{\mu_i}{x_i}
\left[g(x)-g(R_i x)\right]R_i\Psi.
\label{dunkl-gauge-commutator}
\end{equation}
Thus, the Dunkl derivative $D_i$ does not transform covariantly under an arbitrary local gauge transformation. When the gauge function satisfies $g(R_i x)=g(x)$, the additional reflection-dependent term vanishes and
the remaining contribution has the standard derivative form. Gauge covariance can then be implemented by introducing the electromagnetic covariant derivative and transforming the gauge potential accordingly.
For a general gauge function, a generalized gauge-covariant Dunkl connection would be required.

The Coulomb interaction considered in Sec.~6 is a fixed, static, radial external potential. Since $r$ is invariant under every $R_i$, it preserves the discrete reflection sectors. Nevertheless, the present model is not a
dynamical gauge theory and does not claim covariance under arbitrary local gauge transformations.

Under a Lorentz transformation $x'^\mu=\Lambda^\mu{}_\nu x^\nu$, the Klein--Gordon solution transforms as a scalar, $\Psi'(x')=\Psi(x)$~\cite{Greiner2000,Ryder1996}. Introducing the space-time differential operator
$\mathbb{D}_\mu=(\partial_t,D_i)$, ordinary Lorentz covariance would require
\begin{equation}
\mathbb{D}'_\mu\Psi'(x')
=
\Lambda_\mu{}^\nu\mathbb{D}_\nu\Psi(x).
\label{lorentz-covariance-condition}
\end{equation}
This relation is not fulfilled for generic $\mu_i\neq0$, because the operators $R_i$ refer to fixed spatial reflection hyperplanes and are not mixed with the time coordinate by the Dunkl construction
~\cite{Dunkl1989,Rosler2003,HamilLutfuogluMerad2025b}. For the Dunkl wave operator $\Box_D=\partial_t^2-\sum_{i=1}^{d}D_i^2$ and the ordinary Lorentz generators $M_{\mu\nu}=i(x_\mu\partial_\nu-x_\nu\partial_\mu)$, the ordinary Lorentz generators do not commute, in general, with the Dunkl wave operator. For generic boosts and continuous rotations that do not preserve the reflection hyperplanes, one has
\begin{equation}
[M_{0i},\Box_D]\neq0,
\qquad
[M_{ij},\Box_D]\neq0,
\qquad
\mu_i\neq0.
\label{dunkl-lorentz-commutators}
\end{equation}
The second relation applies to generic continuous rotations; only transformations preserving the chosen reflection system remain exact spatial symmetries.

By contrast, the spatial Dunkl Laplacian commutes with every reflection operator~\cite{Dunkl1989,Rosler2003}, as written in the form
\begin{equation}
\left[
\sum_{i=1}^{d}D_i^2,R_j
\right]
=
0,
\qquad
j=1,\ldots,d.
\label{dunkl-reflection-invariance}
\end{equation}
The radial $\mathfrak{su}(1,1)$ generators consequently act within each fixed reflection sector. They provide a dynamical symmetry of the separated radial problem and should not be interpreted as generators of
space-time Lorentz transformations.

The symmetry reduction identified above is explicit: it originates from the fixed reflection hyperplanes and the spatial deformation parameters. It should therefore not be interpreted as spontaneous symmetry breaking.
No quantum anomaly is involved in the present analysis, since the reduced symmetry is already present at the level of the differential--reflection equation. In the non-deformed limit $\mu_i\to0$, the reflection-dependent
terms disappear, $\Box_D\to\Box$, and the ordinary Lorentz and gauge-covariant Klein--Gordon formulation is recovered~\cite{Greiner2000,Ryder1996}.

\subsection{Nonrelativistic Coulomb limit}
As an additional consistency test, we derive the nonrelativistic limit of the Coulomb sector. Although natural units are used throughout the manuscript, $\hbar$ and $c$ are temporarily restored to display explicitly
the expansion in powers of $1/c^2$. The stationary Dunkl--Klein--Gordon equation is written as
\begin{equation}
\left[
\hbar^2c^2\Delta_D
+
(E-V)^2
-
m^2c^4
\right]\Psi
=
0,
\qquad
V(r)
=
-\frac{Ze^2}{r}.
\label{relativistic-dunkl-coulomb-limit}
\end{equation}
For the positive-energy branch, the nonrelativistic energy is introduced as $E=mc^2+\varepsilon$, where $|\varepsilon|,|V|\ll mc^2$. The energy-dependent term then takes the form
\begin{equation}
(E-V)^2-m^2c^4
=
2mc^2(\varepsilon-V)
+
(\varepsilon-V)^2.
\label{nonrelativistic-energy-expansion}
\end{equation}
Substitution of Eq.~\eqref{nonrelativistic-energy-expansion} into
Eq.~\eqref{relativistic-dunkl-coulomb-limit} gives
\begin{equation}
\left[
-\frac{\hbar^2}{2m}\Delta_D
+
V(r)
\right]\Psi
=
\left[
\varepsilon
+
\frac{(\varepsilon-V)^2}{2mc^2}
\right]\Psi.
\label{expanded-dunkl-coulomb-equation}
\end{equation}
The second term on the right-hand side is a relativistic correction of order $1/c^2$. At leading order, Eq.~\eqref{expanded-dunkl-coulomb-equation} therefore reduces to
\begin{equation}
\left[
-\frac{\hbar^2}{2m}\Delta_D
-
\frac{Ze^2}{r}
\right]\Psi
=
\varepsilon\Psi,
\label{dunkl-schrodinger-coulomb-equation}
\end{equation}
which is the $d$-dimensional Dunkl--Schr\"odinger equation with an attractive Coulomb potential.

The spectral reduction can be obtained directly from Eq.~\eqref{energy-coulomb}. Introducing the dimensionless Coulomb coupling $\alpha_Z=Ze^2/(\hbar c)$ and the quantity $\mathcal{A}=2L+\Sigma_\mu+(d-2)/2$, the positive-energy branch takes the form
\begin{equation}
E_{n,L}^{\mathrm C}
=
mc^2
\left[
1+
\frac{\alpha_Z^2}{
\left(
n+\frac{1}{2}
+
\sqrt{\mathcal{A}^2-\alpha_Z^2}
\right)^2
}
\right]^{-1/2}.
\label{dimensional-coulomb-spectrum}
\end{equation}
Since $\alpha_Z=\mathcal{O}(c^{-1})$, the square root in Eq.~\eqref{dimensional-coulomb-spectrum} can be expanded as
\begin{equation*}
\sqrt{\mathcal{A}^2-\alpha_Z^2}
=
|\mathcal{A}|
-
\frac{\alpha_Z^2}{2|\mathcal{A}|}
+
\mathcal{O}(\alpha_Z^4),
\qquad
\mathcal{A}\neq0.
\end{equation*}
Consequently, the expansion of the relativistic spectrum takes the form
\begin{equation}
E_{n,L}^{\mathrm C}
=
mc^2
-
\frac{
mZ^2e^4
}{
2\hbar^2
\left[
n+\frac{1}{2}
+
\left|
2L+\Sigma_\mu+\frac{d-2}{2}
\right|
\right]^2
}
+
\mathcal{O}(c^{-2}).
\label{nonrelativistic-coulomb-spectrum-expansion}
\end{equation}
After writing $E_{n,L}^{\mathrm C}=mc^2+\varepsilon_{n,L}$, the leading nonrelativistic binding energy is therefore given by
\begin{equation}
\varepsilon_{n,L}
=
-
\frac{
mZ^2e^4
}{
2\hbar^2
\left[
n+\frac{1}{2}
+
\left|
2L+\Sigma_\mu+\frac{d-2}{2}
\right|
\right]^2
}.
\label{dunkl-coulomb-binding-energy}
\end{equation}
For the conventional branch $\mathcal{A}>0$, the absolute value reduces to $|\mathcal{A}|=\mathcal{A}$, and the effective principal quantum number in Eq.~\eqref{dunkl-coulomb-binding-energy} becomes $n+2L+\Sigma_\mu+(d-1)/2$. The resulting expression agrees with the known spectrum of the $d$-dimensional Dunkl--Coulomb Schr\"odinger problem ~\cite{Genest2013b,Ghazouani2020,Ghazouani2021}. Thus, both the wave
equation and the relativistic spectrum possess the correct nonrelativistic limit.

\subsection{Non-deformed and three-dimensional limits}
The non-deformed limit is obtained by setting $\mu_i=0$ for all $i$, so that $D_i\to\partial_i$ and $\sum_{i=1}^{d}D_i^2\to\Delta_d$. In this limit, the angular eigenvalue given by Eq.~(\ref{OMEGAL}) takes the form
\begin{equation}
\varpi^2
=
4L\left(L+\frac{d-2}{2}\right)
=
\ell(\ell+d-2).
\label{ordinary-angular-eigenvalue}
\end{equation}
Here, $\ell=2L$ is the conventional orbital quantum number. Equation~\eqref{ordinary-angular-eigenvalue} is the standard eigenvalue of the angular Laplacian on $S^{d-1}$ and agrees with the higher-dimensional analytical result reported in Ref.~\cite{HamilLutfuogluMerad2025b}.

Substitution of Eq.~\eqref{ordinary-angular-eigenvalue} into the radial equations (\ref{EDSEHD}) and (\ref{DIFSECOCP}) recovers the ordinary $d$-dimensional Klein--Gordon equations for the oscillator and Coulomb interactions,
respectively. In particular, the oscillator spectrum in Eq.~(\ref{SPECDK}) can be written as
\begin{equation}
E_{n\ell}^{\mathrm{osc}}
=
\pm\sqrt{m^2+2m\omega(2n+\ell)},
\label{ordinary-oscillator-spectrum}
\end{equation}
in agreement with
Refs.~\cite{BruceMinning1993,HamilLutfuogluMerad2025b}. Similarly, the non-deformed Coulomb spectrum takes the form
\begin{equation}
E_{n\ell}^{\mathrm C}
=
\pm m
\left[
1+
\frac{Z^2e^4}{
\left(
n+\frac{1}{2}
+
\sqrt{
\left(\ell+\frac{d-2}{2}\right)^2-Z^2e^4
}
\right)^2}
\right]^{-1/2}.
\label{ordinary-coulomb-spectrum}
\end{equation}
Equation~\eqref{ordinary-coulomb-spectrum} coincides with the exact $d$-dimensional result derived in Ref.~\cite{Dong2003}. In the notation of that reference, the radial and principal quantum numbers are related by
$N=n_0+\ell+1$, so that $N-\ell-\tfrac12=n_0+\tfrac12$. Therefore, the identification $n=n_0$ establishes the equivalence of the two quantization conventions.

The conventional three-dimensional theory follows by further setting $d=3$. In this case, the angular eigenvalue becomes $\varpi^2=\ell(\ell+1)$, while the angular functions reduce to the spherical harmonics $Y_{\ell m}(\theta,\phi)$. The regular oscillator radial solutions can then be written as
\begin{equation}
R_{n\ell}^{\mathrm{osc}}(r)
=
\mathcal{N}_{n\ell}^{\mathrm{osc}}\,
r^\ell e^{-m\omega r^2/2}
L_n^{\ell+1/2}(m\omega r^2),
\label{ordinary-three-dimensional-oscillator-radial}
\end{equation}
in agreement with the ordinary three-dimensional Klein--Gordon oscillator solutions~\cite{BruceMinning1993}. For the three-dimensional Coulomb problem, the remaining parameter entering the radial solution is introduced as
\begin{equation}
q=\sqrt{\left(\ell+\frac{1}{2}\right)^2-Z^2e^4}.
\label{ordinary-three-dimensional-coulomb-parameters}
\end{equation}
Using this parameter together with $a=\sqrt{m^2-E^2}$, defined in Sec.~6, the regular radial solutions take the form
\begin{equation}
R_{n\ell}^{\mathrm C}(r)
=
\mathcal{N}_{n\ell}^{\mathrm C}\,
e^{-ar}(2ar)^{q-1/2}L_n^{2q}(2ar).
\label{ordinary-three-dimensional-coulomb-radial}
\end{equation}
This radial function and the $d=3$ specialization of Eq.~\eqref{ordinary-coulomb-spectrum} agree with the exact $D$-dimensional result of Ref.~\cite{Dong2003}.

The ordinary rotational symmetry is also recovered in the non-deformed limit. For arbitrary $d$, the generators $L_{ij}=-i(x_i\partial_j-x_j\partial_i)$ span the $\mathfrak{so}(d)$ algebra. In the particular case $d=3$, defining
$L_i=\frac{1}{2}\epsilon_{ijk}L_{jk}$ yields
\begin{equation}
[L_i,L_j]
=
i\epsilon_{ijk}L_k.
\label{ordinary-rotation-algebra}
\end{equation}
These operators generate the $\mathfrak{so}(3)$ algebra and commute with both central Hamiltonians, so that the states can be classified by the usual angular quantum numbers $(\ell,m)$. Thus, the standard radial
solutions, energy spectra, and rotational symmetry are recovered only when $d=3$ and $\mu_i=0$ for every $i$; setting $d=3$ alone does not remove the Dunkl deformation.

Finally, in the non-deformed limit, the Bargmann indices and Sturmian bases reduce to their standard radial $\mathrm{SU}(1,1)$ forms. The corresponding Perelomov coherent states retain the usual representation-theoretical
structure under the algebraic evolution generated by the radial $\mathfrak{su}(1,1)$ operator~\cite{Perelomov1986,GurMann2005}.

\section{Concluding Remarks}
In this work, we have investigated the $d$-dimensional Dunkl--Klein--Gordon equation within an algebraic framework based on the $\mathfrak{su}(1,1)$ symmetry. By employing the Schr\"odinger factorization method, we constructed explicitly the generators of the noncompact Lie algebra and established a unified algebraic treatment for both the Dunkl--Klein--Gordon oscillator and the Dunkl--Coulomb problem in arbitrary spatial dimensions. The corresponding energy spectra and Sturmian radial eigenfunctions were obtained analytically through the theory of irreducible unitary representations.

The analysis shows that the dimensionality and the Dunkl deformation parameters enter the problem through effective radial contributions that modify the localization properties and spectral structure of the system. In the oscillator sector, the reflection operators generate parity-dependent energy shifts and produce a characteristic inversion of the algebraic towers between positive and negative deformation sectors. In the Coulomb sector, the spectrum is organized according to the generalized angular momentum parameter, exhibiting the typical relativistic accumulation of highly excited states near the continuum threshold. Furthermore, the radial probability densities reveal that positive Dunkl deformations tend to broaden the distributions and shift them toward larger radial regions, whereas negative deformations reinforce the localization around the origin.

The algebraic towers displayed for $\mu_i=\pm0.4$ illustrate the spectral effects induced by the Dunkl deformation. Positive deformation shifts the towers toward higher energies, increases their separation across the
reflection sectors, and produces a larger Bargmann index $k$. In contrast, negative deformation shifts the towers toward lower energies. Additional calculations for the intermediate values $\mu_i=\pm0.2$ and $\mu_i=\pm0.3$
exhibit the same qualitative trends, with the spectral displacement and the corresponding variation of $k$ becoming progressively more pronounced as $|\mu_i|$ increases.

The Coulomb towers displayed for $\mu_i=\pm0.4$ illustrate how the Dunkl deformation modifies the spectral organization associated with $L=0,1,2$. For positive deformation, the Bargmann index $k_L^{\mathrm C}$ increases and the towers remain within the real positive-discrete representation. For negative deformation, physical admissibility depends more sensitively on the combined effects of $\Sigma_\mu$, $L$, and the Coulomb coupling. Additional calculations, not displayed, show that the $L=0$ sector becomes nonphysical for $\mu_i=-0.2$ because its Bargmann index is complex, whereas the more negative value $\mu_i=-0.4$ again yields a real index. This apparently
nonmonotonic behavior follows from the quadratic dependence of the reality condition on the signed combination $2L+\Sigma_\mu+(d-2)/2$.

The oscillator and Coulomb interactions also exhibit distinct spectral organizations. The oscillator has an equally spaced ladder in the $\mathfrak{su}(1,1)$ representation variable, although the corresponding relativistic energies are not generally equally spaced. By contrast, the Coulomb bound-state energies become progressively compressed and accumulate near the continuum threshold as the radial quantum number increases.
For the parameter sets explicitly displayed in the oscillator figures, the corresponding sectors retain real spectra and positive-discrete representations. 

More generally, however, sufficiently negative deformations combined with increasing dimensionality may lead to nonphysical oscillator sectors with $E^{2}<0$, as discussed below. In the Coulomb sector, certain negative deformations may instead produce complex Bargmann indices and exclude specific angular-momentum towers. Spectral admissibility is therefore controlled jointly by the Dunkl deformation, angular momentum, interaction type, and Coulomb coupling.

As the dimensionality $d$ increases, the effective Dunkl contribution $\sum_i \mu_i=d\mu$ significantly modifies the structure of the spectrum. In the oscillator sector, for negative deformations, certain reflection sectors may generate complex energies when the Dunkl contribution dominates over the positive term $2(n+L)$, particularly in the fully negative sector and for sufficiently large values of $d$ with fixed angular momentum $L$. Nevertheless, increasing the angular momentum restores the condition $E^2>0$, leading again to a completely real oscillator spectrum. This behavior shows that the spectral stability of the oscillator sector is governed by the interplay between the dimensionality, the Dunkl deformation, the reflection sector, and the angular momentum.

Within the same algebraic construction, radial Perelomov coherent states were obtained for both interactions. Their physically normalized probability densities were analyzed with respect to the Dunkl radial measure. Under the evolution generated by the corresponding radial $\mathfrak{su}(1,1)$ operator, the coherent parameter undergoes a phase rotation that produces a periodic breathing of the radial packet. This algebraic evolution should be distinguished from the exact physical Klein--Gordon evolution, particularly in the Coulomb sector, where the relativistic energies depend nonlinearly on the radial quantum number. The dimensionality and the Dunkl parameters determine the radial region and spatial extension of the coherent profiles.

From a physical perspective, the arbitrary-dimensional formalism makes it possible to distinguish the effects of dimensionality, angular momentum, and
Dunkl deformation. The results show that the reflection structure reorganizes the spectral sectors and modifies the radial localization, while the spatial
dimension introduces an additional geometric contribution. In the oscillator sector, the reflections differentiate the parity sectors; in the Coulomb
problem, the interplay among deformation, angular momentum, and coupling determines the stability of the representations and the existence of physical
bound-state towers.

The non-deformed limit recovers the ordinary $d$-dimensional Klein--Gordon equations, spectra, and rotational structure. Its three-dimensional specialization reproduces the conventional oscillator and Coulomb radial solutions together with the $\mathfrak{so}(3)$ symmetry. Moreover, the positive-energy Coulomb branch reduces to the $d$-dimensional Dunkl--Schr\"odinger equation and its known binding spectrum in the nonrelativistic limit. These reductions provide direct consistency checks of the present formalism against established results~\cite{Ghazouani2021,Mota2021A,Hamil2022, HamilLutfuogluMerad2025b}.

The principal contribution of the present study is therefore not merely the recovery of the oscillator and Coulomb spectra, which are consistent with previous differential treatments, but their derivation and organization
within a unified higher-dimensional $\mathfrak{su}(1,1)$ framework. The explicit construction of the generators, positive-discrete representations, Sturmian bases, and Perelomov coherent states provides an algebraic
description that complements the exact solutions previously reported for lower and arbitrary dimensions \cite{Mota2021A,Hamil2022,HamilLutfuogluMerad2025b,Ghazouani2021}. Within this formulation, the ladder structure and the algebraic bound-state sectors follow from a common representation-theoretical construction, while their physical admissibility is established by the regularity, normalization, and bound-state conditions derived in Sec.~7.

\appendix
\renewcommand{\theequation}{A.\arabic{equation}}
\setcounter{equation}{0}

\section{Operator conventions for the $\mathfrak{su}(1,1)$ algebra}
The Lie algebra $\mathfrak{su}(1,1)$ provides an algebraic framework in which the spectral properties of a quantum system can be described through representation theory. In the convention adopted here, its generators
$\mathcal{R}_1$, $\mathcal{R}_2$, and $\mathcal{R}_3$ satisfy the commutation relations~\cite{Vourdas1990}
\begin{equation}
[\mathcal{R}_1,\mathcal{R}_2]
=
-i\mathcal{R}_3,
\qquad
[\mathcal{R}_2,\mathcal{R}_3]
=
i\mathcal{R}_1,
\qquad
[\mathcal{R}_3,\mathcal{R}_1]
=
i\mathcal{R}_2.
\label{su11-cartesian-commutators}
\end{equation}
Defining the ladder operators by $\mathcal{R}_{\pm}=\mathcal{R}_1\pm i\mathcal{R}_2$, the previous relations can equivalently be expressed as
\begin{equation}
[\mathcal{R}_3,\mathcal{R}_{\pm}]
=
\pm\mathcal{R}_{\pm},
\qquad
[\mathcal{R}_{-},\mathcal{R}_{+}]
=
2\mathcal{R}_3,
\qquad
\mathcal{R}_3^\dagger=\mathcal{R}_3,
\qquad
\mathcal{R}_{+}^{\dagger}=\mathcal{R}_{-}.
\label{su11-ladder-commutators}
\end{equation}
The quadratic Casimir operator is invariant under the action of the algebra and therefore characterizes its irreducible representations. In terms of both sets of generators, it is written as
\begin{equation}
\mathcal{C}^{2}
=
\mathcal{R}_3^{2}
-
\mathcal{R}_1^{2}
-
\mathcal{R}_2^{2}
=
\mathcal{R}_3^{2}
-
\frac{1}{2}
\left(
\mathcal{R}_{+}\mathcal{R}_{-}
+
\mathcal{R}_{-}\mathcal{R}_{+}
\right),
\qquad
[\mathcal{C}^{2},\mathcal{R}_i]=0.
\label{su11-casimir}
\end{equation}
Among the unitary representations of $\mathfrak{su}(1,1)$, the bound-state solutions considered in this manuscript belong to the positive discrete series. Its basis is denoted by
$\{|k,n\rangle;\,n=0,1,2,\ldots\}$, where $k>0$ is the Bargmann index. The action of the generators on these states takes the form
\begin{align}
\mathcal{R}_{+}|k,n\rangle
&=
\sqrt{(n+1)(2k+n)}\,|k,n+1\rangle,
\nonumber\\
\mathcal{R}_{-}|k,n\rangle
&=
\sqrt{n(2k+n-1)}\,|k,n-1\rangle,
\qquad
\mathcal{R}_3|k,n\rangle
=
(k+n)|k,n\rangle.
\label{su11-generator-action}
\end{align}
The state $|k,0\rangle$ is the normalized lowest-weight state and satisfies $\mathcal{R}_{-}|k,0\rangle=0$. The Casimir operator acts on every state of the representation according to
$\mathcal{C}^{2}|k,n\rangle=k(k-1)|k,n\rangle$.

In the oscillator sector, these conventions are implemented through the identification
$(\mathcal{R}_3,\mathcal{R}_{\pm})=(K_0,Q_{\pm})$.
For the Coulomb sector, the corresponding identification is
$(\mathcal{R}_3,\mathcal{R}_{\pm})
=(\mathcal{H}_0,\mathcal{G}_{\pm})$.
Thus, both operator realizations satisfy the same
$\mathfrak{su}(1,1)$ commutation relations.
\section*{Acknowledgments}
B. C. L. is grateful to the Excellence project FoS UHK 2205/2025-2026 for the financial support. This work was partially supported by SNI--M\'exico, EDI--IPN, and SIP--IPN under project number 20260918.

\section*{Data Availability Statement}
No new data were generated or analyzed in this study.

\section*{ORCID iDs}
M. Salazar-Ram\'irez: \url{https://orcid.org/0000-0003-4139-3026}\\
B. C. L\"utf\"uo\u{g}lu: \url{https://orcid.org/0000-0001-6467-5005}\\

\end{document}